\documentclass[fleqn,10pt]{wlscirep}

\newcommand{\Vfire}{V_{\mathsf{T}}} 

\newcommand{\Vsat}{V_{\mathsf{S}}}  

\newcommand{\Vreset}{V_{\mathsf{R}}} 

\newcommand{\Vbase}{V_{\mathsf{B}}}  

\newcommand{\Wbif}{W_{\mathsf{B}}} 

\newcommand{\Wcrit}{W_{\mathsf{C}}} 

\newcommand{\gcrit}{\Gamma_{\mathsf{C}}}  

\newcommand{\gbif}{\Gamma_{\mathsf{B}}}  

\newcommand{\rhocrit}{\rho_{\mathsf{C}}}  

\newcommand{\set}[1]{\left\{#1\right\}} % Set with explicit elements.
\newcommand{\ludremoved}[1] {}

\title{Phase transitions and self-organized criticality 
in networks of stochastic spiking neurons}

\author[1]{Ludmila Brochini}
\author[2]{Ariadne de Andrade Costa}
\author[1]{Miguel Abadi}
\author[3]{Antônio C. Roque}
\author[2]{Jorge Stolfi}
\author[3,*]{Osame Kinouchi}

\affil[1]{Universidade de São Paulo, Departamento de Estatística-IME, 
 São Paulo-SP, 05508-090, Brazil}
\affil[2]{Universidade de Campinas, Instituto de Computação, 
 Campinas-SP, 13083-852, Brazil}
\affil[3]{Universidade de São Paulo, Departamento de Física-FFCLRP, 
 Ribeirão Preto-SP, 14040-901, Brazil} 
 
\affil[*]{okinouchi@gmail.com}
\begin{abstract}
Phase transitions and critical behavior are
crucial issues both in theoretical and experimental 
neuroscience. We report analytic and 
computational results about phase transitions and self-organized
criticality (SOC) in
networks with general stochastic neurons. The stochastic neuron 
has a firing probability given by a smooth monotonic function $\Phi(V)$
of the membrane potential $V$, rather than a sharp firing threshold.
We find that such networks can operate in several
dynamic regimes (phases) depending on the average synaptic weight
and the shape of the firing function $\Phi$.  
In particular, we encounter both continuous 
and discontinuous phase transitions to absorbing states. 
At the continuous transition critical boundary, neuronal 
avalanches occur whose distributions of size and duration are given by 
power laws, as observed in biological neural networks.
We also propose and test a new mechanism to produce SOC: 
the use of dynamic neuronal gains -- a form of
short-term plasticity probably in the axon initial segment (AIS) -- 
instead of depressing synapses at the dendrites 
(as previously studied in the literature). 
The new self-organization mechanism produces a slightly supercritical
state, that we called SOSC, 
in accord to some intuitions of Alan Turing.
\end{abstract}

\begin{document}

\flushbottom
\maketitle

\thispagestyle{empty}

\quote{\emph{Another simile would be an atomic pile of less than critical size: an injected idea is to correspond to a neutron entering the pile from without. Each such neutron will cause a certain disturbance which eventually dies away. If, however, the size of the pile is sufficiently increased, the disturbance caused by such an incoming neutron will very likely go on and on increasing until the whole pile is destroyed. Is there a corresponding phenomenon for minds, and is there one for machines? There does seem to be one for the human mind. The majority of them seems to be \emph{subcritical}, i.e., to correspond in this analogy to piles of subcritical size. An idea presented to such a mind will on average give rise to less than one idea in reply. A smallish proportion are supercritical. An idea presented to such a mind may give rise to a whole "theory" consisting of secondary, tertiary and more remote ideas. (...) Adhering to this analogy we ask, "Can a machine be made to be supercritical?"} Alan Turing (1950)\cite{Turing1950}.}

\section*{Introduction}

The Critical Brain Hypothesis ~\cite{Chialvo2010,Hesse2015} states that (some) 
biological neuronal networks work near phase transitions because criticality 
enhances information processing capabilities~\cite{Kinouchi2006,Beggs2008,Shew2009}  
and health ~\cite{Massobrio2015}. The first discussion about criticality 
in the brain, in the sense that subcritical, critical and slightly 
supercritical branching process of thoughts could describe human 
and animal minds, has been made in the beautiful speculative 1950 Imitation Game paper by 
Turing~\cite{Turing1950}. In 1995, Herz {\&} Hopfield~\cite{Herz1995} noticed 
that self-organized criticality (SOC) models for earthquakes were 
mathematically equivalent to networks of integrate-and-fire neurons, 
and speculated that perhaps SOC would occur in the brain. In 2003, 
in a landmark paper, these theoretical conjectures found experimental 
support by Beggs and Plenz ~\cite{Beggs2003} and, by now, more than 
half a thousand papers can be found about the subject, 
see some reviews~\cite{Chialvo2010,Markovic2014,Hesse2015}. 
Although not consensual, the Critical Brain Hypothesis can be 
considered at least a very fertile idea.

The open question about neuronal criticality is what are the mechanisms 
responsible for tuning the network towards the critical state. 
Up to now, the main mechanism studied is some dynamics in the links which, 
in the biological context, would occur at the synaptic
level~\cite{Arcangelis2006,Levina2007,
Bonachela2010,Arcangelis2012,Costa2015,Kessenich2016,Campos2016}. 

Here we propose a whole new mechanism: dynamic neuronal gains, related to 
the diminution (and recovery) of the firing probability, an intrinsic 
neuronal property. 
The neuronal gain is experimentally related to the well known phenomenon 
of firing rate adaptation~\cite{Ermentrout2001,Benda2003,Buonocore2016}. 
This new mechanism is sufficient to drive neuronal networks of stochastic 
neurons towards a critical boundary found, by the first time, for these models. 
The neuron model we use was proposed by Galves and Locherbach~\cite{Galves2013} 
as a stochastic model of spiking neurons
inspired by the traditional integrate-and-fire (IF) model. 

Introduced in the early 20th century~\cite{Lapicque1907}, 
IF elements have been extensively used in simulations of spiking 
neurons~\cite{Gerstein1964,Burkitt2006a,Burkitt2006b,Naud2012,
Brette2007,Brette2015,Buonocore2016}. 
Despite their simplicity, IF models have successfully emulated 
certain phenomena observed in biological neural networks, such as firing
avalanches~\cite{Levina2007,Bonachela2010,Benayoun2010}
and multiple dynamical regimes~\cite{Ostojic2014,Torres2015}.
In these models, the membrane potential $V(t)$ integrates synaptic 
and external currents up to a \emph{firing threshold} 
$\Vfire$~\cite{Platkiewicz2010}. 
Then, a spike is generated and $V(t)$ drops to
a \emph{reset potential} $\Vreset$. The
\emph{leaky integrate-and-fire} (LIF) model extends the IF neuron
with a leakage current, which causes the potential $V(t)$ to 
decay exponentially towards a \emph{baseline potential} $\Vbase$ 
in the absence of input signals~\cite{Burkitt2006a,Naud2012}.  

LIF models are deterministic but  
it has been claimed that stochastic models may be more adequate 
for simulation purposes~\cite{Mcdonnell2016}.  
Some authors proposed to introduce stochasticity by 
adding noise terms to the potential~\cite{Burkitt2006a,Burkitt2006b,
Ostojic2014,Torres2015,Mcdonnell2016,Soula2006,
Cessac2008,Cessac2010,Cessac2011}, yielding the 
\emph{leaky stochastic integrate-and-fire} (LSIF) models. 

 Alternatively, the Galves-Löcherbach (GL) 
 model ~\cite{Galves2013,DeMasi2015,Duarte2014,Duarte2015,Galves2016} 
 and also the model used by Larremore \emph{et al.}~\cite{Larremore2014, Virkar2016}
 introduce stochasticity in their firing neuron models in 
 a different way. Instead of noise inputs, they 
assume that the firing of the neuron is a random event, 
whose probability of occurrence in any time 
step is a \emph{firing function} $\Phi(V)$ of membrane 
potential $V$. By subsuming all sources of randomness into
a single function, the Galves-Löcherbach (GL) neuron model 
simplifies the analysis and simulation of noisy spiking neural networks. 

Brain networks are also known to exhibit \emph{plasticity}:
changes in neural parameters over time scales longer 
than the firing time scale~\cite{Brette2007,Cooper2005}.   
For example, short-term synaptic plasticity~\cite{Tsodyks1998} 
has been incorporated in models by assuming that the 
strength of each synapse is lowered after 
each firing, and then gradually recovers towards a reference
value~\cite{Levina2007,Bonachela2010}.
This kind of dynamics drives the synaptic weights of the 
network towards critical values, a SOC state
which is believed to optimize the network information
processing~\cite{Beggs2003,Kinouchi2006,
Larremore2011,Markovic2014,Hesse2015,Massobrio2015}.
  
In this work, first we study the dynamics of networks of GL 
neurons by a very simple and transparent mean-field calculation.
We find both continuous and discontinuous phase transitions
depending on the average synaptic strength and parameters of the firing 
function $\Phi(V)$. To the best of our knowledge, these phase transitions
have never been observed in standard integrate-and-fire neurons. 
We also find that, at the second order 
phase transition the stimulated excitation of a single 
neuron causes avalanches of firing events (neuronal avalanches)  that 
are similar to those observed in biological
networks~\cite{Beggs2003,Hesse2015}.

Second, we present a new mechanism for SOC 
based on a dynamics on the \emph{neuronal gains} (a parameter
of the neuron probably related to the axon initial segment 
-- AIS~\cite{Kole2012,Platkiewicz2010}), 
instead of depression of coupling strengths (related to 
neurotransmiter vesicle depletion at synaptic contacts between neurons)
proposed in the literature~\cite{Levina2007,Bonachela2010,Costa2015,Campos2016}. 
This new activity dependent gain model 
is sufficient to achieve self-organized criticality, 
both by simulation evidence and by mean-field calculations.
The great advantage of this new SOC mechanism is that it
is much more efficient, since we have only one adaptive parameter 
per neuron, instead of one per synapse.

\section*{The Model}\label{s.model}

We assume a network of $N$ GL neurons that 
change states in parallel at certain \emph{sampling times} 
with a uniform spacing $\Delta$. 
Thus, the membrane potential of neuron $i$ is modeled by a real variable
$V_i[t]$ indexed by \emph{discrete time} $t$, an integer that 
represents the sampling time $t\Delta$.

Each synapse transmits signals from some \emph{presynaptic} neuron $j$ to 
some \emph{postsynaptic} neuron $i$, 
and has a \emph{synaptic strength} $w_{ij}$.
If neuron $j$ fires between discrete 
times $t$ and $t+1$, its potential drops to $\Vreset$. 
This event increments by $w_{ij}$ the potential of every postsynaptic 
neuron $i$ that does not fire in that interval. 
The potential of a non-firing neuron
may also integrate an \emph{external stimulus} $I_i[t]$, 
which can model signals received from sources outside
the network. Apart from these increments, 
the potential of a non-firing neuron decays at each 
time step towards the baseline voltage $\Vbase$ by a 
factor $\mu \in [0,1]$, which models 
the effect of a leakage current. 

We introduce the Boolean variable $X_i[t]\in \set{0,1}$ which 
denotes whether neuron $i$ fired between $t$ and $t+1$.
The potentials evolve as:
\begin{equation}
V_i[t+1] = 
  \left\{
     \begin{array}{lcl}
        \displaystyle
        \Vreset  &\quad& \hbox{if $X_i[t]= 1$,} \\
        \displaystyle
         \mu (V_i[t]-\Vbase) + \Vbase + I_i[t] + 
         \sum_{j=1}^{N} w_{i j} X_j[t] &\quad&
            \hbox{if $X_i[t] = 0$.}
     \end{array}
   \right.
\label{modeldiscrete}
\end{equation}
This is a special case of the general GL model~\cite{Galves2013}, 
with the filter function $g(t-t_s) = \mu^{t - t_s}$, 
where $t_s$ is the time of the last firing of neuron $i$.
We have $X_i[t+1]=1$ with probability $\Phi(V_i[t])$, which is called the 
\emph{firing function}~\cite{Galves2013,Larremore2014,
DeMasi2015,Duarte2014,Duarte2015,Galves2016}.
We also have $X_i[t+1]=0$ if $X_i[t]=1$ (refractory period).
The function $\Phi$ is sigmoidal, that is, monotonically 
increasing, with limiting values $\Phi(-\infty)= 0$ and 
$\Phi(+\infty) = 1$, with only one derivative
maximum. We also assume that $\Phi(V)$ is 
zero up to some \emph{threshold potential} $\Vfire$ (possibly
$-\infty$) and is 1 starting at some \emph{saturation potential} 
$\Vsat$ (possibly $+\infty$).
If $\Phi$ is the shifted Heaviside step function 
$\Theta$, $\Phi(V) = \Theta(V - \Vfire)$, 
we have a deterministic discrete-time LIF neuron. 
Any other choice for $\Phi(V)$ gives a stochastic neuron.

The network's activity is measured by the fraction (or density)
$\rho[t]$ of firing neurons:
\begin{equation}
  \rho[t] = \frac{1}{N} \sum_{j=1}^N X_j[t]\:.
  \label{e.rhot}
\end{equation}
The density $\rho[t]$ can be computed from the 
probability density $p[t](V)$ of potentials at time $t$:
\begin{eqnarray}
  \rho[t] = \int_{\Vfire}^\infty \Phi(V) p[t](V)\, dV \:,
  \label{e.rho0}
\end{eqnarray}
where $p[t](V)\,dV$ is the fraction of neurons with potential 
in the range $[V,V+dV]$ at time $t$.

Neurons that fire between $t$ and $t+1$ have their 
potential reset to $\Vreset$.  
They contribute to $p[t+1](V)$ a Dirac impulse at 
potential $\Vreset$, with amplitude 
(integral) $\rho[t]$ given by equation~(\ref{e.rho0}).
In subsequent time steps, the potentials of all neurons
will evolve according to equation~(\ref{modeldiscrete}). 
This process modifies $p[t](V)$ also for $V \neq \Vreset$.

\section*{Results}

We will study only fully connected networks, where
each neuron receives inputs from all the other $N-1$ neurons.
Since the zero of potential is arbitrary, we assume $\Vbase = 0$.
We also consider only the case with $\Vreset = 0$,
and uniform constant input $I_i[t]=I$. So, for these networks, 
equation~(\ref{modeldiscrete}) reads:
\begin{equation}
V_i[t+1] = 
  \left\{
     \begin{array}{lcl}
        \displaystyle
        0  &\quad& \hbox{if $X_i[t]= 1$,} \\
        \displaystyle
         \mu V_i[t]  + I +
         \sum_{j=1}^{N} w_{ij} X_j[t] &\quad&
            \hbox{if $X_i[t] = 0$.}
     \end{array}
   \right.
\label{model}
\end{equation}

\subsection*{Mean-field calculation}

In the mean-field analysis, we assume that the synaptic 
weights $w_{ij}$ follow a distribution with average $W/N$ 
and finite variance.  
The mean-field approximation disregards correlations, 
so the final term of equation~(1) becomes:
\begin{equation}
\sum_{j=1}^N w_{ij} X_j[t] = W \rho[t] \:.
\end{equation}
Notice that the variance of the weights $w_{ij}$ 
becomes immaterial when $N$ tends to infinity. 

Since the external input $I$ is the same for all neurons and all times, 
every neuron $i$ that does not fire between $t$ 
and $t+1$ (that is, with $X_i[t]= 0$) has its
potential changed in the same way: 
\begin{equation}
V_i[t+1] = \mu V_i[t] + I + W \rho[t]\:,
\label{e.rhoinc}
\end{equation}
Recall that the probability density $p[t](V)$ has a Dirac impulse 
at potential $U_0=0$, 
representing all neurons that fired in the previous interval.  
This Dirac impulse is modified in later 
steps by equation~(\ref{e.rhoinc}).
It follows that, once all neurons have fired at least once, 
the density $p[t](V)$ will be a combination 
of discrete impulses with amplitudes 
$\eta_0[t],\eta_1[t], \eta_2[t], \ldots$, 
at potentials $U_0[t], U_1[t], U_2[t], \ldots$,
such that $\sum_{k=0}^\infty \eta_k\;=\;1$. 

The amplitude $\eta_k[t]$ is the fraction 
of neurons with \emph{firing age} $k$ 
at discrete time $t$, that is, neurons that fired between
times $t-k-1$ and $t-k$, and did not fire between $t-k$ and $t$.  
The common potential of those neurons, at time $t$, is $U_k[t]$.  
In particular, $\eta_0[t]$ is the fraction 
$\rho[t-1]$ of neurons that fired in the 
previous time step. For this type of 
distribution, the integral of equation~(\ref{e.rho0}) 
becomes a discrete sum:
\begin{equation}
\rho[t]= \sum_{k=0}^{\infty} \Phi(U_k[t]) \: \eta_k[t]\:.
\label{e.rhosum}
\end{equation}
According to equation~(\ref{e.rhoinc}), 
the values $\eta_k[t]$ and $U_k[t]$ evolve by the equations
\begin{eqnarray}
\eta_k[t+1] &=& \left( 1 - \Phi(U_{k-1}[t])\right)\label{e.eta}\:
\eta_{k-1}[t] \:,\\
U_k[t+1] &=& \mu U_{k-1}[t] + I +  W \rho[t]\:,
\label{e.tvk}
\end{eqnarray}
for all $k\geq 1$, with $\eta_0[t+1] = \rho[t]$ and $U_0[t+1] = 0$.

\subsection*{Stationary states for general $\Phi$ and $\mu$}

A \emph{stationary state} is a density $p[t](V)=p(V)$ 
of membrane potentials that does not change with time.
In such a regime, quantities $U_k$ and 
$\eta_k$  do not depend anymore on $t$. 
Therefore, the equations~(\ref{e.eta}--\ref{e.tvk}) 
become the recurrence equations:
\begin{eqnarray}
 \rho = \eta_0 = \sum_{k=0}^{\infty} \Phi(U_k) \eta_k\:,\label{e.seta0}\\
 U_0 = 0 \:,\\
\eta_k &=& \left( 1 - \Phi(U_{k-1})\right)\eta_{k-1} \:,\label{e.srhok}\\
U_k &=& \mu U_{k-1} + I +  W \rho\:,
\label{e.svk}
\end{eqnarray}
for all $k\geq 1$.  

\begin{figure}[!ht]
\centering  % figura centralizada
\includegraphics[width=0.7\linewidth]{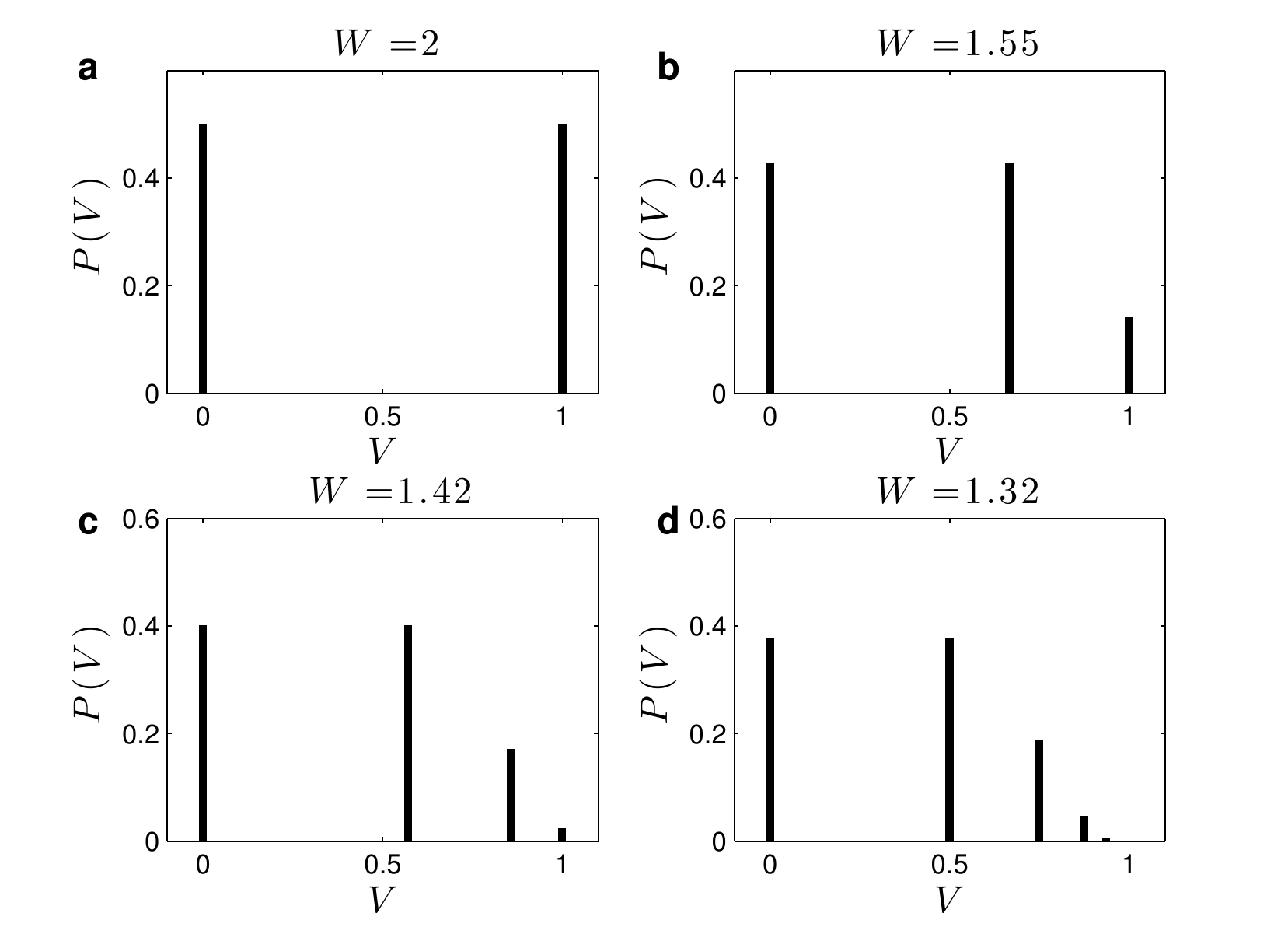}
 \caption{{\bf Examples of stationary potential 
distributions $P(V)$:} monomial $\Phi$ function with
$r=1,\Gamma =1,\mu = 1/2, I=0$ case with different 
values of $W$. a) $W_2=\Wbif=2$, two peaks;
b) $W_3=14/9$, three peaks; c) $W_4=488/343$, four peaks, 
d) $W_\infty \approx 1.32$, infinite number of peaks with $U_\infty=1$. 
Notice that for $W < W_\infty$ all the peaks in the distribution $P(V)$ 
lie at potentials $U_k < 1$. For $\Wbif=2$ we have 
$\eta_0=\eta_1=1/2$, producing a bifurcation to a \emph{2-cycle}.
The values of $W_m = W_2, W_3, W_4$ and $W_\infty$ can be obtained 
analytically by imposing the condition $U_m=1$ in equations 
(\ref{e.srhok}--\ref{e.svk}).}
\label{peaks}
\end{figure}

Since equations~(\ref{e.srhok}) 
are homogeneous on the $\eta_k$, the normalization 
condition $\sum_{k=0}^\infty \eta_k\;=\;1$ must be included explicitly. 
So, integrating over the density $p(V)$ leads to a 
discrete distribution $P(V)$ (see Fig.~\ref{peaks} for a specific
$\Phi$).

Equations~(\ref{e.seta0}--\ref{e.svk})
can be solved numerically, e.~g.~by simulating the evolution of 
the potential probability density $p[t](V)$ according to 
equation~(\ref{e.eta}--\ref{e.tvk}), starting 
from an arbitrary initial distribution, until reaching a stable
distribution (the probabilities $\eta_k$ should be 
renormalized for unit sum after each time step, 
to compensate for rounding errors). Notice that this can be done for
any $\Phi$ function, so this numerical solution is very general.

\subsection*{The monomial saturating $\Phi$ with $\mu > 0$}

Now we consider a specific class of firing functions, 
the \emph{saturating monomials}. 
This class is parametrized by a positive \emph{degree} $r$ and a 
\emph{neuronal gain} $\Gamma > 0$. 
In all functions of this class, $\Phi(V)$ is $0$ when 
$V\leq \Vfire$, and $1$ when 
$V \geq \Vsat$, where the saturation potential is
$\Vsat = \Vfire+ 1/\Gamma$. In the interval $\Vfire < V < \Vsat$,
we have:
\begin{equation}
  \Phi(V) = \left(\Gamma (V-\Vfire)^{\vphantom{0}}\right)^r \:. 
  \label{e.linphi}
\end{equation}
Note that these functions can be 
seen as limiting cases of sigmoidal functions, and that we recover 
the deterministic LIF model 
$\Phi(V) = \Theta(V-\Vfire)$ when $\Gamma \rightarrow \infty$.

For any integer $p \geq 2$, there are combinations of values of $\Vfire$,
$\Vsat$, and $\mu$ that cause the network to behave deterministically.
This happens if the stationary state defined by equations~(\ref{e.srhok})
and~(\ref{e.svk}) is such that $U_{p-2}\leq \Vfire \leq \Vsat \leq
U_{p-1}$---that is, $\Phi(U_k)$ is either 0 or 1 for all $k$, so the GL
model becomes equivalent to the deterministic LIF model.  In such a
stationary state, we have $\rho = \eta_k=1/p$ for all $k< p$; meaning that
the neurons are divided into $p$ groups of equal size, and each group
fires every $p$ steps, exactly. If the inequalities are strict ($U_{p-2} <
\Vfire$ and $\Vsat < U_{p-1}$) then there are also many deterministic
periodic regimes (\emph{$p$-cycles}) where the $p$ groups have slightly
more or less than $1/p$ of all the neurons, but still fire regularly every
$p$ steps.

Note that, if $\Vfire = 0$, such degenerate (deterministic) regimes,
stationary or periodic, occur only for $p= 2$ and $W \geq \Wbif$ where
$\Wbif = 2(I + \Vsat)$. The stationary regime has $\rho = \eta_0 = \eta_1
= 1/2$ and $U_1 = I + W/2$. In the periodic regimes (\emph{2-cycles}) the
activity $\rho[t]$ alternates between two values $\rho'$ and $\rho'' =
1-\rho'$, with $\rho_1(W) < \rho' < 1/2 < \rho'' < \rho_2(W)$, where:
\begin{equation}
  \rho_1(W) = \frac{\Vsat}{W}
  \quad\quad\hbox{and}\quad\quad
  \rho_2(W) = 1 - \rho_1(W) = \frac{W - \Vsat}{W}
  \label{eqbounds}
\end{equation}
All these \emph{2-cycles} are marginally stable, in the sense that,
if a perturbed state $\rho_\epsilon=\rho+\epsilon$ satisfy 
equation~(\ref{eqbounds}) then the new cycle $\rho_\epsilon[t+1]=
1-\rho_\epsilon[t]$ is also marginally stable.

\begin{figure}[!ht]
  \centering  % figura centralizada
 \makebox[\textwidth][c]{ \includegraphics[width=18cm]
 {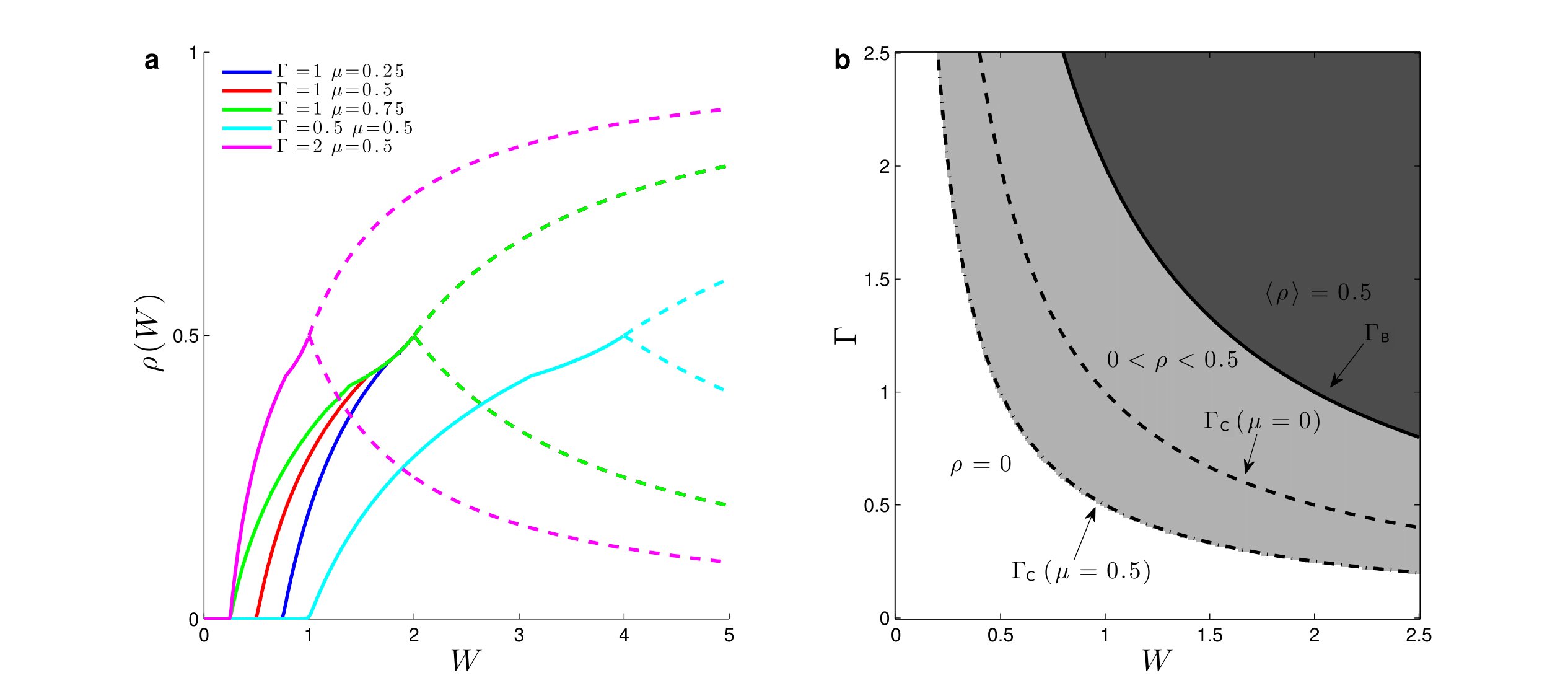}}
  \caption{{\bf Results for $\mu > 0$ :} a) Numerically computed 
  $\rho(W)$ curves for the monomial $\Phi$ with $r = 1$, 
  $I = \Vreset = \Vfire = 0$, and  
 $(\Gamma,\mu)=(1,1/4), (1,1/2), (1,3/4), (1/2,1/2)$, and $(2,1/2)$.
 The absorbing state $\rho_0$ looses stability at $\Wcrit$ and the
 non trivial fixed point $\rho>0$ appears. At $\Wbif = 2/\Gamma$, we 
 have $\rho=1/2$ and from there we have the fixed point 
 $\rho[t] = 1/2$ and the \emph{2-cycles} with $\rho[t]$ 
 between the two bounds of equation~(\ref{eqbounds}) (dashed lines).
 b) Numerically computed $(\Gamma,W)$ diagram 
 showing the critical boundaries $\gcrit(W) = (1-\mu)/W$ and the bifurcation 
 line $\gbif(W) = 2/W$ to \emph{2-cycles}.}
\label{mupos}
\end{figure}

In the analyses that follows, the control parameters 
are $W$ and $\Gamma$, and $\rho(W,\Gamma)$ is the order parameter.
We obtain numerically $\rho(W,\Gamma)$
and the phase diagram $(W,\Gamma)$
for several values of $\mu > 0$, for the linear ($r=1$) saturating
$\Phi$ with $I =\Vfire = 0$ (Fig.~\ref{mupos}).  
Only the first 100 peaks $(U_k,\eta_k)$ were considered,
since, for the given $\mu$ and $\Phi$, there was no significant 
probability density beyond that point. The same numerical method
can be used for $r \neq 1, I \neq 0, \Vfire \neq 0$.

Near the critical point, we obtain numerically
$\rho(W,\mu) \approx C (W-\Wcrit)/W $, where 
$\Wcrit(\Gamma)= (1-\mu)/\Gamma$ and $C(\mu)$ is a constant.
So, the critical exponent is $\alpha = 1$,
characteristic of the mean-field directed percolation 
(DP) universality class~\cite{Kinouchi2006,Hesse2015}.
The critical boundary in the $(W,\Gamma)$ plane, numerically obtained,
seems to be $\gcrit(W) = (1-\mu)/W$  (Fig.~\ref{mupos}b). 

\subsection*{Analytic results for $\mu = 0$}

Below we give results of a simple mean-field analysis 
in the limits $N\rightarrow \infty$ and $\mu\rightarrow 0$.  
The latter implies that, at time $t+1$, the neuron ``forgets'' 
its previous potential $V_i[t]$ and 
integrates only the inputs $I[t] + W_{ij} X_j[t]$. 
This scenario is interesting because it enables analytic solutions, 
yet exhibits all kinds of behaviors and
phase transitions that occur with $\mu > 0$.

When $\mu = 0$ and $I_i[t]=I$ (uniform constant input), 
the density $p[t](V)$ consists of only
two Dirac peaks at potentials $U_0[t] = \Vreset = 0$ and
$U_1[t]=  I + W \rho[t-1]$, with fractions 
$\eta_0[t]$ and $\eta_1[t]$ that evolve as:
\begin{eqnarray}
  \eta_0[t+1]  &=&  \rho[t] = \Phi(0)\eta_0[t] 
  + \Phi(I+W\eta_0[t])(1-\eta_0[t])\:, \label{e.rho0-mu0}\\
  \eta_1[t+1] & =& 1 - \eta_0[t+1]\:. \label{e.rho1-mu0}
\end{eqnarray} 
Furthermore, if the neurons cannot fire spontaneously, that is,
$\Phi(0) = 0$, then equation~(\ref{e.rho0-mu0}) reduces to:
\begin{equation}
  \eta_0[t+1] = \rho[t] = \Phi(I + W\eta_0[t])(1 - \eta_0[t]) \:. 
  \label{e.rho21}
\end{equation}
In a stationary regime, equation~(\ref{e.rho21}) simplifies to:
\begin{equation}
  \rho =  (1-\rho) \Phi(I + W \rho) \:, \label{e.rho20}
\end{equation}
since $\eta_0 = \rho$, $\eta_1 = 1 - \rho$, 
$U_0= 0$, and $U_1 = I+W \rho$. Below, all the results refer to
the monomial saturating $\Phi$s given by equation~(\ref{e.linphi}).

\subsubsection*{The case with $r = 1, \Vfire=0$}

\begin{figure}[!ht]
 \centering  % figura centralizada
\makebox[\textwidth][c]{ \includegraphics[width=\linewidth]{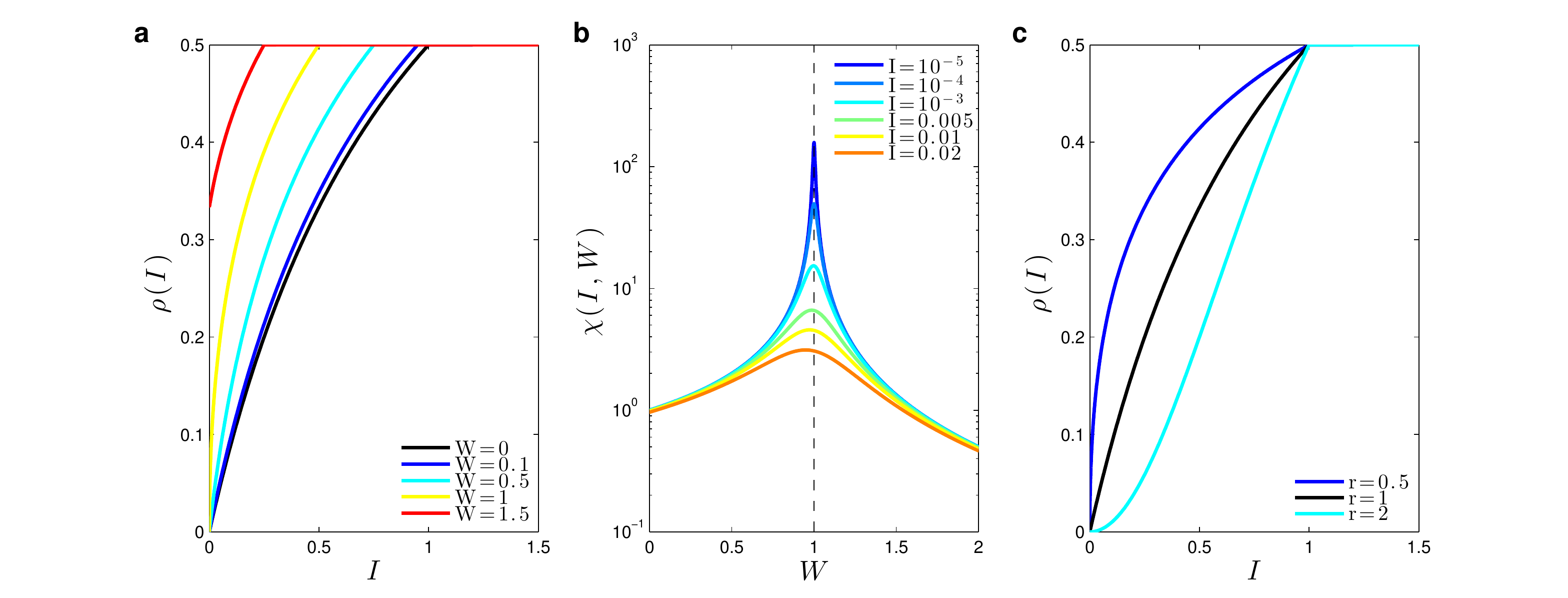}}
 \caption{{\bf Network and isolated neuron responses to external input $I$:}
 a) Network activity $\rho(I,W)$ as a function of $I$ for several $W$;
 b) Susceptibility $\chi(I,W)$ as a function of $W$ for several $I$.
 Notice the divergence $\chi_{\mathsf{C}}(I) 
 \propto I^{-1/2}$ for small $I$;
 c) Firing rate of an isolated neuron $\rho(I,W=0)$ for 
 monomial exponents $r=0.5,1$ and $2$. }
\label{FI}
\end{figure}

When $r = 1$, we have the linear function $\Phi(V)= \Gamma V $ for 
$0<V<\Vsat = 1/\Gamma$, where $V = I + W \rho$. Equation~(\ref{e.rho20})
turns out:
\begin{equation}
\Gamma  W \rho^2 - (\Gamma W - \Gamma I - 1)\rho - \Gamma I = 0\:,
\end{equation}
with solution (Fig.~\ref{FI}a):
\begin{equation}
\rho = \frac{\Gamma W - \Gamma I - 1 + \sqrt{(\Gamma W - \Gamma I - 1)^2
+ 4 \Gamma^2 W I}}{2 \Gamma W}\:. \label{rhoWI}
\end{equation}

For zero input we have:
\begin{equation}
\rho(W) = \frac{(W-\Wcrit)^\beta}{W} \:, \label{rhocrit}
\end{equation}
where $\Wcrit = 1/\Gamma$ and the order parameter critical 
exponent is $\beta = 1$. This corresponds to a standard mean-field
continuous (second order) absorbing state phase transition. This
transition will be studied in detail two section below.

A measure of the network sensitivity to inputs (which play here the role
of external fields) is the susceptibility $\chi = d\rho/dI$, 
which is a function of $\Gamma, W$ and $I$ (Fig.~\ref{FI}b):
\begin{equation}
\chi  = \frac{\Gamma(1-\rho)}
{2\Gamma W \rho -\Gamma W +\Gamma I+1} \:.
\end{equation}
For zero external inputs, the susceptibility behaves as:
\begin{equation}
\chi(W) = \frac{1}{\Gamma W} (W - \Wcrit)^{-\gamma} \:,
\end{equation}
where we have the critical exponent $\gamma = 1$.

A very interesting result is that, for any $I$, the susceptibility
is maximized at the critical line $\Wcrit = 1/\Gamma$, with the values:
\begin{eqnarray}
\rhocrit &=& \frac{-\Gamma I + \sqrt{\Gamma^2 I^2 + 4 \Gamma I} }{2}\:,\\
\chi_{\mathsf{C}}  &=& \frac{\Gamma (2+\Gamma I - 
\sqrt{\Gamma^2 I^2+4 \Gamma I}) }
{2 \sqrt{\Gamma^2 I^2+4\Gamma I}}  \:,
\end{eqnarray}
For $I\rightarrow 0$ we have $\rhocrit \propto \sqrt{I}$.
The critical exponent $\delta$ is defined by $I \propto \rho^\delta$
for small $I$, so we obtain the 
mean-field value $\delta = 2$. In analogy with
Psychophysics, we may call $m = 1/\delta = 1/2$ the Stevens's
exponent of the network~\cite{Kinouchi2006}.

With two critical exponents it is possible to 
obtain others through scaling relations. For example,
notice that $\beta,\gamma$ and $\delta$ are related to
$2\beta+\gamma = \beta (\delta + 1)$.

Notice that, at the critical line, the susceptibility diverges as 
$\chi_{\mathsf{C}} \propto 1/\sqrt{I}$ as $I \rightarrow 0$. 
We will comment the importance of the fractionary Stevens's
exponent $m=1/2$ (Figs.~\ref{FI}a) and the diverging susceptibility
(Figs.~\ref{FI}b) for information processing in the Discussion section.  

\subsubsection*{Isolated neurons}

We can also analyze the behavior of the GL neuron model 
under the standard experiment where an isolated neuron in 
vitro is artificially injected with a current 
of constant intensity $J$.  That corresponds to setting the external input 
signal $I[t]$ of that neuron to a constant value $I = J\Delta/C$ where 
$C$ is the effective capacitance of the neuron.

The firing rate of an isolated neuron can be written as:
 \begin{equation}
 F(I) = \rho(I) F_{\max}\:;
 \end{equation}
where $F_{\max}$ is an empirical maximum firing rate (measured in
spikes per second) of a given neuron and $\rho$ is our 
previous neuron firing probability per time step. With $W=0$ and 
$I > 0$ in equation~(\ref{e.rho20}), we get:
\begin{equation}
  \rho(I) = \Phi(I)\left(1-\rho(I)\right) \:, \label{e.rho.iso}
\end{equation}
The solution for the monomial saturating $\Phi$ with $\Vfire=0$ 
is (Fig.~\ref{FI}c):
\begin{equation}
\rho(I) = \frac{(\Gamma I)^r}{1+(\Gamma I)^r}\:,
\label{e.rho.isolin}
\end{equation}
which is less than $\rho = 1/2$ only if $I <  1/\Gamma$.
For any $I \geq 1/\Gamma$ the firing rate saturates at $\rho = 1/2$ 
(the neuron fires at every other step, alternating between 
potentials $U_0 = \Vreset = 0$ and $U_1 = I$.
So, for $I>0$, there is no phase transition.  
Interestingly, equation~(\ref{e.rho.isolin}), known as 
generalized Michaelis-Menten function, 
is frequently used to fit the firing response of biological 
neurons to DC currents~\cite{Lipetz1971,Naka1966}.

\subsubsection*{Continuous phase transitions in networks: 
the case with $r=1$}

\begin{figure}[!ht]
  \makebox[\textwidth][c]{\includegraphics[width=0.75\linewidth]{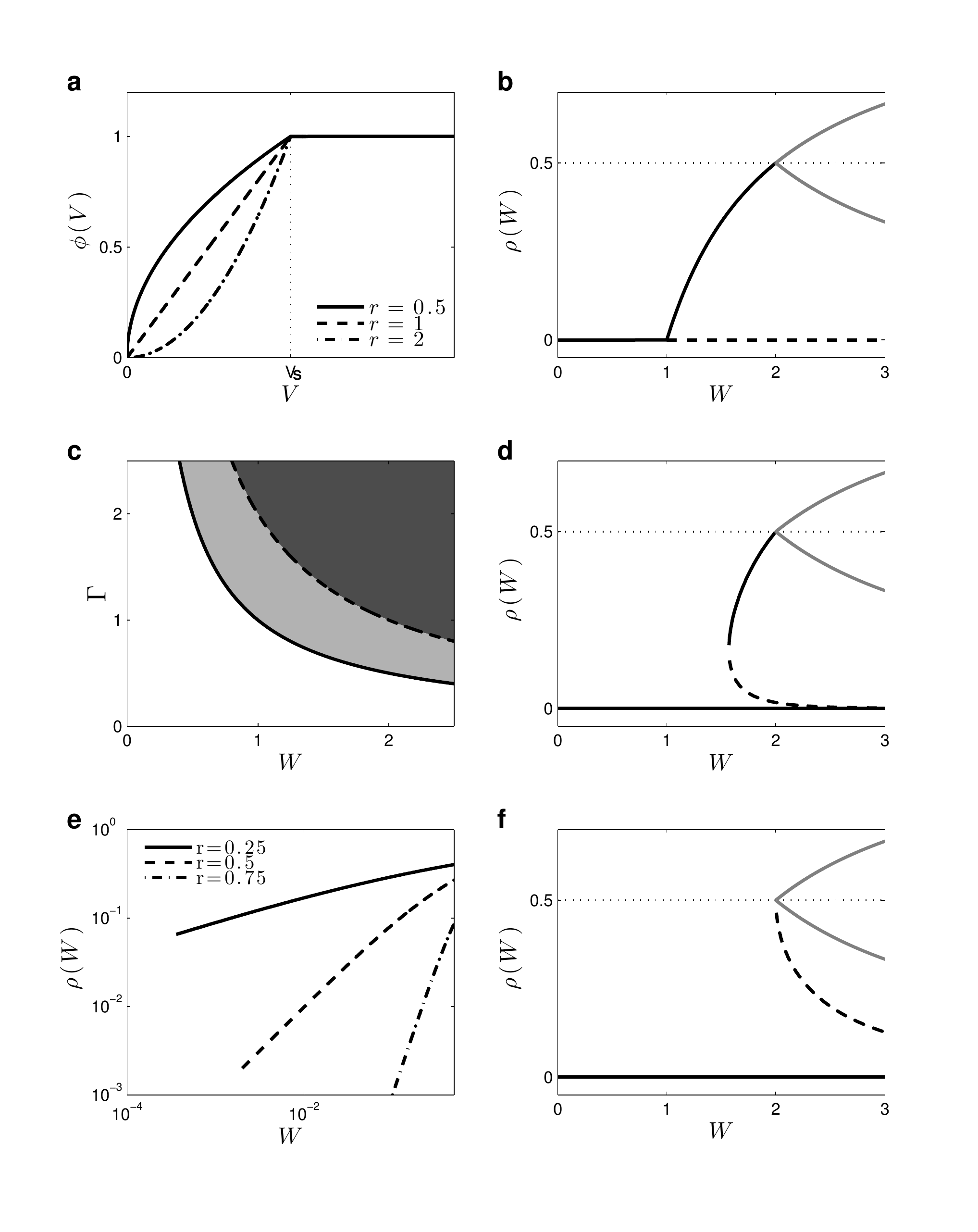}}
    \caption{{
    \bf Firing densities (with $\Gamma = 1$)
    and phase diagram with $\mu=0$ and $\Vfire = 0$.} 
  a) Examples of monomial firing functions $\Phi(V)$ with $\Gamma=1$
  $r = 0.5, 1$ and $2$. 
  b) The $\rho(W)$ bifurcation plot for $r=1$.
  The absorbing state $\rho_0$ looses stability after $W>\Wcrit=1$
  (dashed line).
  The non trivial fixed point $\rho^+$ bifurcates at $\Wbif = 2/\Gamma = 2$ 
  into two branches (gray lines) that bound the 
  marginally stable \emph{2-cycles}. 
  c) The $(\Gamma, W)$ phase diagram for $r=1$.  
  Below the critical boundary $\Gamma = \gcrit(W) = 1/W$ 
  the inactive state $\rho=0$ is absorbing and stable; 
  above that line it is also absorbing but unstable.
  Above the line $\Gamma = \gbif(W) = 2/W$ there
  are only the marginally stable \emph{2-cycles}. For 
  $\gcrit(W) < \Gamma \leq \gbif(W)$ there is
  a single stationary regime $\rho(W) = (W-\Wcrit)/W < 1/2$,
  with $\Wcrit = 1/\Gamma$. 
  d) Discontinuous phase transitions 
  for $\Gamma=1$ with exponents $r=1.2$. The absorbing state 
  $\rho_0$ now is stable (solid line at zero). The non trivial
  fixed point $\rho^+$ starts with the value $\rhocrit$ 
  at $\Wcrit$ and bifurcates at $\Wbif$, creating the boundary
  curves (gray) that delimit possible \emph{2-cycles}. At $\Wcrit$ 
  also appears the unstable separatrix $\rho_-$ (dashed line).
  e) Ceaseless activity (no phase transitions) for 
  $r = 0.25, 0.5$ and $r=0.75$. The activity approach zero
  (for $W=0$ as power laws. f) In the limiting case $r=2$ we
  do not have a $\rho > 0$ fixed point, but only
  the stable $\rho = 0$ (black), the \emph{2-cycles} region (gray)
  and the unstable separatrix (traces). 
  }
\label{mu0r}
\end{figure}

Even with $I = 0$, spontaneous collective activity is possible if
the network suffers a phase transition. 
With $r=1$, the stationary state condition equation~(\ref{e.rho20}) is:
\begin{equation}
  \Gamma W \rho^2 + (1-\Gamma W) \rho \;=\; 0 \:.
\end{equation}
The two solutions are the absorbing state $\rho=0$ and the non-trivial 
state:
\begin{equation}
  \rho = \frac{W-\Wcrit}{W}\:, 
  \label{e.rholin}
\end{equation}
with $\Wcrit = 1/\Gamma$.
Since we must have $0 < \rho \leq 1/2$, this solution is valid  only for
$\Wcrit < W \leq \Wbif=2/\Gamma$ (Fig~\ref{mu0r}b).  

This solution describes a stationary state where $1-\rho$ of the 
neurons are at potential $U_1 =  W - \Wcrit$. The neurons 
that will fire in the next step are a fraction $\Phi(U_1)$ of those, 
which are again a fraction $\rho$ of the total.  For any $W > \Wcrit$, 
the state $\rho = 0$ is unstable: any small perturbation of the 
potentials cause the network to converge to 
the active stationary state above.
For $W < \Wcrit$, the solution $\rho = 0$ is stable and absorbing. 
In the $\rho(W)$ plot, the locus of stationary regimes 
defined by equation~(\ref{e.rholin}) bifurcates at $W = \Wbif$ 
into the two bounds of equation~(\ref{eqbounds}) 
that delimit the \emph{2-cycles} (Fig.~\ref{mu0r}b). 

So, at the critical boundary $W = 1/\Gamma$, we have a standard continuous 
absorbing state transition $\rho(W) \propto (W-\Wcrit)^\alpha$ 
with a critical exponent $\alpha = 1$, which also can be written
as $\rho(\Gamma) \propto (\Gamma -\gcrit)^\alpha$.
In the $(\Gamma,W)$ plane, the phase transition corresponds 
to a critical boundary $\gcrit(W) = 1/W$, below 
the \emph{2-cycle} phase transition $ \gbif(W) = 2/W$ 
(Fig.~\ref{mu0r}c). 

\subsubsection*{Discontinuous phase transitions in networks: the case with
$r>1$}

When $r > 1$ and $W \leq \Wbif=2/\Gamma$, the 
stationary state condition is:
\begin{equation}
  (\Gamma W)^r \rho^r - (\Gamma W)^r\rho^{r-1}+ 1 = 0\:.
  \label{e.rho.rne1}
\end{equation}
This equation has a non trivial solution $\rho^+$ only when $1 \leq r \leq 2$ 
and $\Wcrit(r) \leq W \leq \Wbif$, for a certain $\Wcrit(r) > 1/\Gamma$. 
In this case, at $W = \Wcrit(r)$, there is a discontinuous 
(first-order) phase transition to a regime with 
activity $\rho = \rhocrit(r) \leq 1/2$ (Fig.~\ref{mu0r}d). 
It turns out that $\rhocrit(r) \rightarrow 0$ as $r \rightarrow 1$,
recovering the continuous phase transition in that limit.
For $r = 2$, the solution to equation~(\ref{e.rho.rne1}) 
is a single point $\rho(\Wcrit) = \rhocrit = 1/2$ 
at $\Wcrit = 2/\Gamma =\Wbif$ (Fig.~\ref{mu0r}f). 

Notice that, in the linear case,
the fixed point $\rho_0=\rho =0$ is unstable for $W>1$ (Fig.~\ref{mu0r}b).
This occurs because the \emph{separatrix} $\rho_-$ (trace lines, 
Fig.~\ref{mu0r}d), for $r \rightarrow 1$,
collapses with the $\rho_0$ point, so that it looses its stability.

\subsubsection*{Ceaseless activity: the case with $r < 1$}

When $r < 1$, there is no absorbing solution $\rho=0$ 
to equation~(\ref{e.rho.rne1}).
In the $W \rightarrow 0$ limit we get 
$\rho(W) = (\Gamma W)^{r/(1-r)} $. These power laws means that
$\rho > 0$  for any $W > \Wcrit(r)=0$ (Fig.~\ref{mu0r}e). We recover the
second order transition $\Wcrit(r=1) = 1/\Gamma$ when $r \rightarrow 1$
in equation~(\ref{e.rho.rne1}). Interestingly, this ceaseless 
activity $\rho>0$ for any $W > 0$ seems to be similar to that found by
Larremore \emph{et al.}~\cite{Larremore2014} with a $\mu=0$ 
linear saturating model. This ceaseless activity, even with $r=1$, 
perhaps is due to the presence of inhibitory neurons in 
Larremore \emph{et al.} model.

\subsubsection*{Discontinuous phase transitions in networks: 
the case with $\Vfire > 0$ and $I > 0$}

The standard IF model has $\Vfire > 0$. If we allow this feature
in our models we find a new ingredient that produces
first order phase transitions. Indeed, in this case, 
if $U_1 = W\rho + I < \Vfire$ then we have a single peak at
$U_0=0$ with $\eta_0=1$, which means we have a silent state.
When $U_1 = W \rho +I > \Vfire$, we have a peak with height
$\eta_1 = 1-\rho$ and $\rho = \eta_0 = \Phi(U_1)\eta_1$.

\begin{figure}[!ht]
  \centering  % figura centralizada
 \makebox[\textwidth][c]{ \includegraphics[width=\linewidth]
 {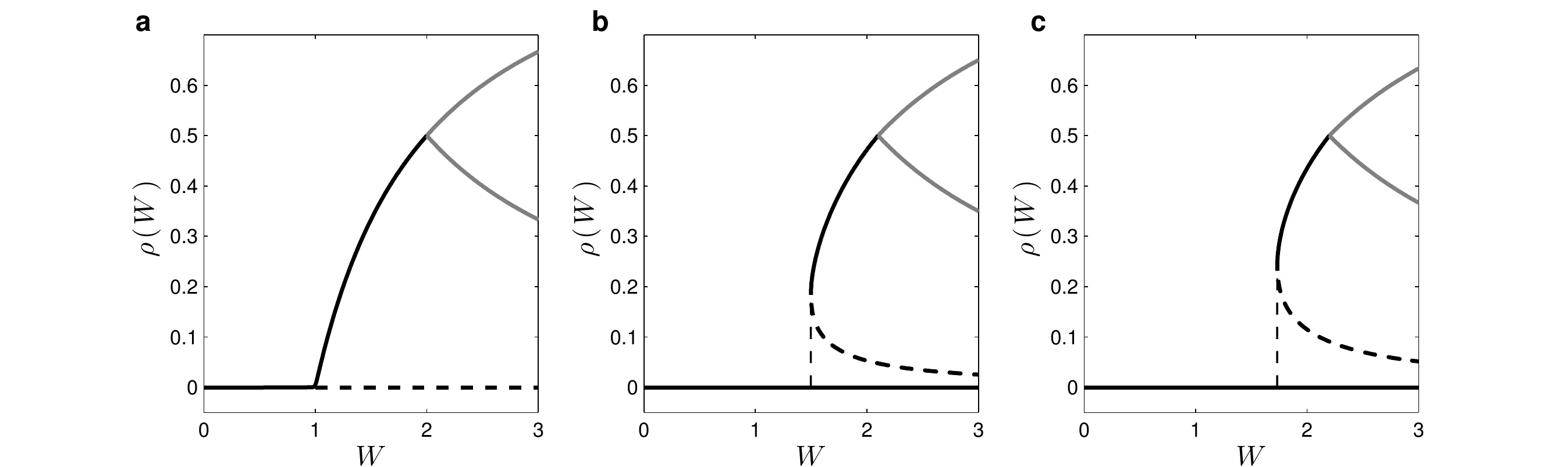}}
 \caption{{\bf Phase transitions for $\Vfire > 0$:} 
 monomial model with $\mu=0$, $r = 1, \Gamma=1$ 
 and thresholds $\Vfire=0, 0.05$ and $0.1$. Here the solid black lines represent the stable fixed points, dashed black lines represent unstable fixed points and grey lines correspond to the marginally stable boundaries of cycles-2 regime.   
 The discontinuity $\rhocrit$ goes to zero for $\Vfire \rightarrow 0$.}   
\label{Vfire}
\end{figure}

For the linear monomial model this leads to the equations:
\begin{eqnarray}
\rho = \Gamma (U_1-\Vfire)(1-\rho)\:,\\
 \Gamma W \rho^2+(1-\Gamma W - \Gamma \Vfire + \Gamma I)\rho 
+\Gamma \Vfire - \Gamma I = 0\:,
\end{eqnarray}
with the solution:
\begin{equation}
\rho^\pm(\Gamma,W,\Vfire,I) = \frac{(\Gamma W + \Gamma \Vfire -\Gamma I -1)
\pm \sqrt{(\Gamma W+\Gamma \Vfire-\Gamma I -1)^2-
4\Gamma^2 W \Vfire+4\Gamma^2WI}}{2\Gamma W} \label{rhoVfire}\:,
\end{equation}
where $\rho^+$ is the non trivial fixed point and $\rho^-$ is the
unstable fixed point (separatrix).
These solutions only exist for $\Gamma W$ values such that
$\Gamma (W + \Vfire -I) - 1 > 2\Gamma \sqrt{W (\Vfire-I)}$.
This produces the condition:
\begin{equation}
\Gamma W > \gcrit \Wcrit = \left(1+\sqrt{\Gamma (\Vfire-I)}\right)^2\:,
\end{equation}
which defines a first order critical boundary.
At the critical boundary the density of firing neurons is:
\begin{equation}
\rhocrit = \frac{\sqrt{\Gamma (\Vfire-I)}}{1+\sqrt{\Gamma (\Vfire-I)}}\:,
\label{rhocVfire}
\end{equation}
which is nonzero (discontinuous) for any $\Vfire >I$.
These transitions can be seen in Fig.~\ref{Vfire}.
The solutions for equations~(\ref{rhoVfire}) and (\ref{rhocVfire}) 
is valid only for $\rhocrit < 1/2$ (\emph{2-cycle} bifurcation). 
This imply the maximal value $\Vfire=1/\Gamma+I$.

\subsection*{Neuronal avalanches}

\begin{figure}[!ht]
  \centering
  \makebox[\textwidth][c]{\includegraphics[width=\linewidth]
  {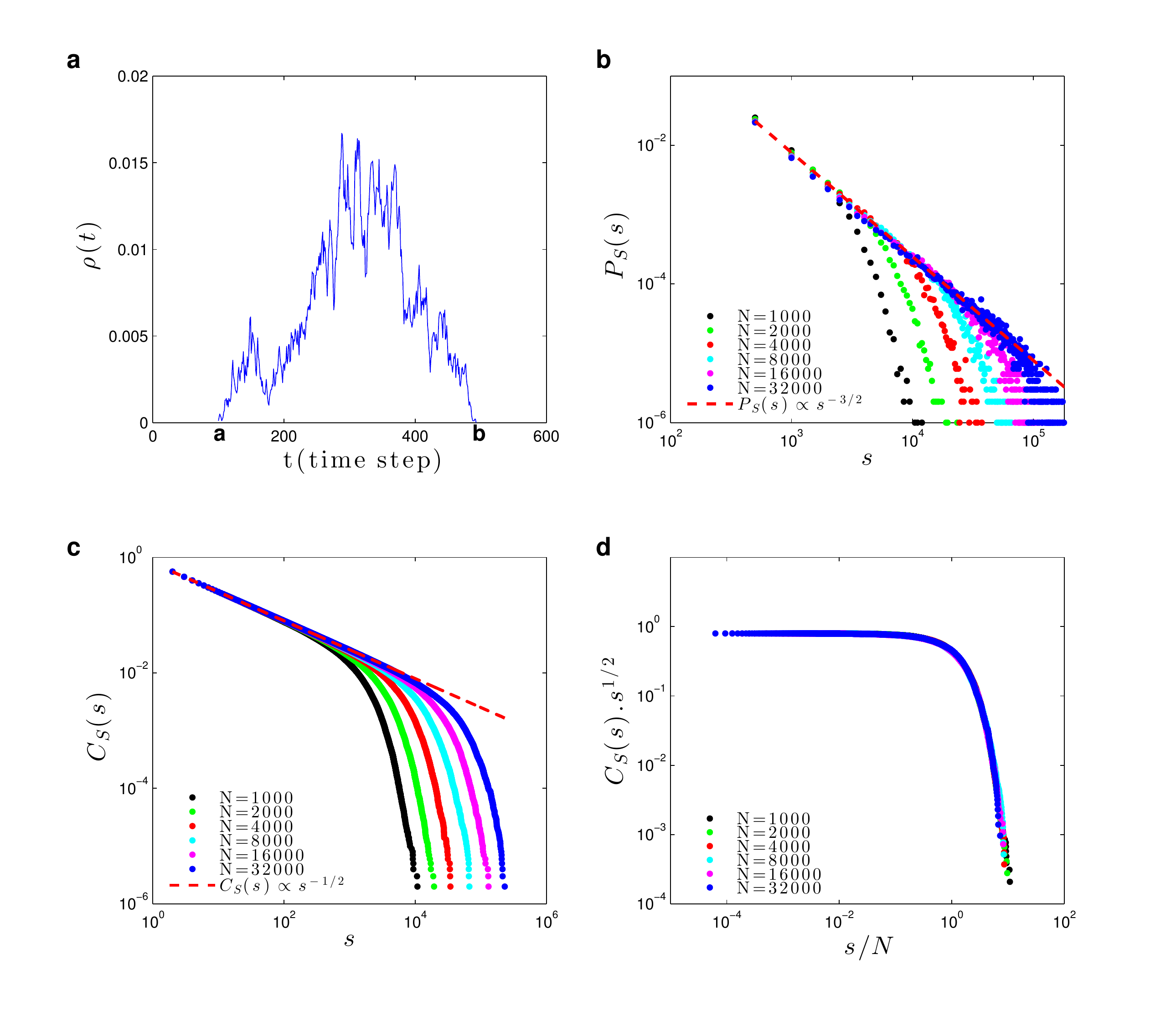}}
  \caption{{\bf Avalanche size statistics in the static model:}
   Simulations at the critical point $\Wcrit=1, \gcrit=1$ (with $\mu=0$ ).
  a) Example of avalanche profile $\rho[t]$ at the critical point.
  b) Avalanche size distribution $P_S(s)\equiv P(S=s)$, for network sizes 
  $N=1000, 2000, 4000, 8000, 16000$ and $32000$. 
  The dashed reference line is proportional 
  to $s^{-\tau_s}$, with $\tau_s = {3/2}$.
  c) Complementary cumulative distribution 
  $C_S(s)= \sum_{k=s}^\infty P_S(k)$. 
  Being an integral of $P_S(s)$, its power law exponent 
  is $-\tau_s+1 = -1/2$ (dashed line).
  d) Data collapse (finite-size scaling) for $C_S(s)s^{1/2}$ versus
function of $s/N^{c_S}$, with the cutoff exponent $c_S =1$.}
\label{Avalanches}
\end{figure}

Firing avalanches in neural networks have attracted significant 
interest because of their possible connection to efficient 
information processing~\cite{Beggs2003,Kinouchi2006,Beggs2008,
Hesse2015,Massobrio2015}. 
Through simulations, we studied the critical 
point $\Wcrit=1,\gcrit=1$ (with $\mu=0$) in search for
neuronal avalanches~\cite{Beggs2003,Hesse2015} (Fig~\ref{Avalanches}). 

An avalanche that starts at discrete time $t=a$ and ends at $t=b$ 
has duration $d = b-a$ and size $s = N \sum_{t=a}^b \rho[t]$
(Fig.~\ref{Avalanches}a).
By using the notation $S$ for a random variable and $s$ for its
numerical value, we observe a power law avalanche size distribution 
$P_S(s) \equiv P(S=s) \propto s^{-\tau_S}$, with the
mean-field exponent $\tau_S=3/2$ (Fig.~\ref{Avalanches}b)~\cite{Beggs2003,
Bonachela2010,Hesse2015}. 
Since the distribution $P_S(s)$ is noisy for large $s$, for further
analysis we use the complementary cumulative function 
$C_S(s) \equiv P(S \geq s) = \sum_{k=s}^{\infty}P_S(k)$ (which
gives the probability of having an avalanche with size equal 
or greater than $s$) because it is very smooth and monotonic
(Fig.~\ref{Avalanches}c). Data collapse gives a
finite-size scaling exponent $c_S=1$ 
(Fig.~\ref{Avalanches}d) \cite{Costa2015,Campos2016}.

\begin{figure}[!ht]
  \centering  % figura centralizada
 \makebox[\textwidth][c]{ \includegraphics[width=\linewidth]
 {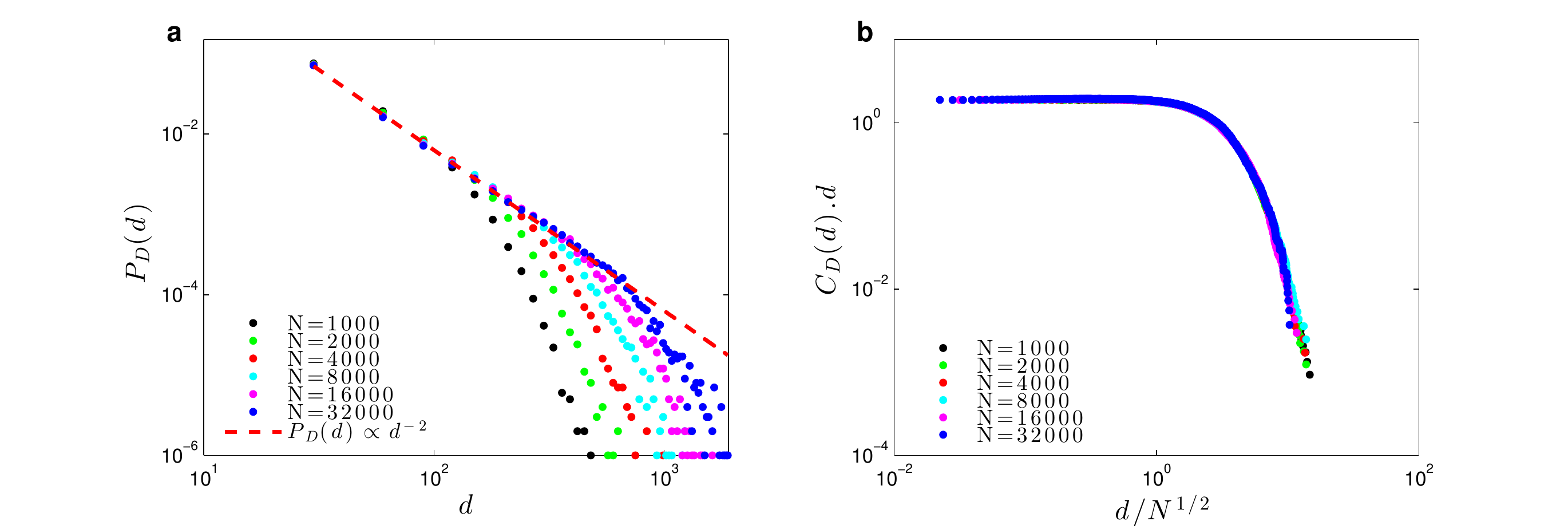}}
  \caption{{\bf Avalanche duration statistics in the static model:} 
  Simulations at the critical point $\Wcrit=1, \gcrit=1$ 
  ($\mu=0$ ) for network sizes 
  $N = 1000, 2000, 4000, 8000, 16000$ and $32000$: 
  a) Probability distribution $P_D(d)\equiv P(D=d)$ 
  for avalanche duration $d$.
  The dashed reference line is proportional to $d^{-\tau_D}$, 
  with $\tau_D = 2$.
  b) Data collapse $C_D(d)d$ versus $d/N^{c_D}$, 
  with the cutoff exponent $c_D = 1/2$. The complementary
  cumulative function $C_D(d) \equiv \sum_{k=d}^\infty P_D(k)$,
  being an integral of $P_D(d)$, has power law exponent
  $-\tau_D + 1 = - 1$.
 }
\label{Duration}
\end{figure}

We also observed a power law distribution for avalanche duration,
$P_D(d) \equiv P(D=d) \propto d^{-\tau_D}$ with $\tau_D=2$
(Figure~\ref{Duration}a).
The complementary cumulative distribution is 
$C_D(d) \equiv P(D\geq d)= \sum_{k=d}^\infty P_D(k)$.
From data collapse, we find a finite-size scaling exponent 
$c_D=1/2$ (Fig.~\ref{Duration}b), 
in accord with the literature~\cite{Bonachela2010}.

\subsection*{The model with dynamic parameters} \label{SOCgain}

The results of the previous section were obtained by fine-tuning the 
network at the critical point $\gcrit=\Wcrit=1$. 
Given the conjecture that the critical region presents functional 
advantages, a biological model should include some homeostatic mechanism
capable of tuning the network towards criticality. Without such mechanism,
we cannot truly say that the network \emph{self-organizes} toward the
critical regime.

\begin{figure}[!ht]
  \centering
  \makebox[\textwidth][c]{\includegraphics[width=\linewidth]
  {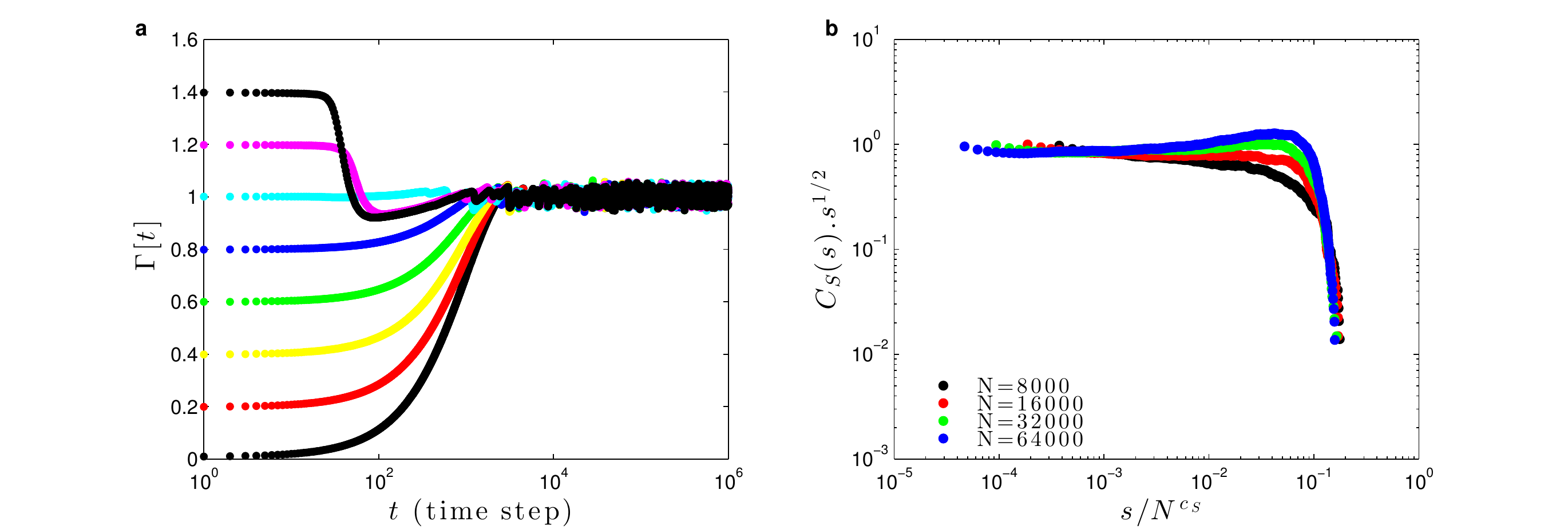}}
  \caption{{\bf Self-organization with dynamic neuronal gains:}
  Simulations of a network of GL neurons with fixed $W_{ij}=W = 1, 
  u=1, A=1.1$ and $\tau = 1000$ ms. Dynamic 
  gains $\Gamma_i[t]$ starts with $\Gamma_i[0]$ uniformly 
  distributed in $[0,\Gamma_{\max}]$. The average initial condition is
  $\Gamma[t] \equiv  \frac{1}{N} \sum_i^N \Gamma_i[t] \approx \Gamma_{\max}/2$,
  which produces the different initial conditions $\Gamma[0]$.
  (a) Self-organization of the average 
  gain $\Gamma[t]$ over time. The horizontal 
  dashed line marks the value $\gcrit=1$. 
  (b) Data collapse for $C_S(s)s^{1/2}$ versus $s/N^{c_S} $ 
  for several $N$, with the cutoff exponent $c_S = 1$.} 
\label{SOC}
\end{figure}

However, observing that the relevant parameter for 
criticality in our model is the critical boundary $\gcrit \Wcrit=1$, 
we propose to work with dynamic gains $\Gamma_i[t]$ while keeping 
the synapses $W_{ij}$ fixed. The idea is to reduce the gain $\Gamma_i[t]$ 
when the neuron fires, and let the gain slowly recover
towards a higher resting value after that:
\begin{equation}
 \Gamma_i[t+1] = \Gamma_i[t] +\frac{1}{\tau}\left(A - \Gamma_i[t]\right) 
 - u \Gamma_i[t] X_i[t]\:.
\label{gt}
\end{equation} 
Now, the factor $\tau$ is related to the 
characteristic recovery time of the gain, $A$ is the asymptotic resting 
gain, and  $u \in [0,1]$ is the fraction 
of gain lost due to the firing.  
This model is plausible biologically, and can be related to a 
decrease and recovery, due to the neuron activity,
of the firing probability at the AIS~\cite{Kole2012}.
Our dynamic $\Gamma_i[t]$ mimics the well known phenomenon 
of \emph{spike frequency adaptation}~\cite{Ermentrout2001,Benda2003}.

Fig.~\ref{SOC}a shows a simulation with all-to-all coupled
networks with $N$ neurons and, for simplicity, $W_{ij}=W$. 
We observe that the average gain 
$\Gamma[t] = \frac{1}{N} \sum_{i=1}^N \Gamma_i[t]$ seems to 
converge toward the critical value $\gcrit(W)=1/W=1$, starting from 
different $\Gamma[0] \neq 1$.  
As the network converges to the critical region,
we observe power-law avalanche size distributions with exponent $-3/2$
leading to a cumulative function $C_S(s) \propto s^{-1/2}$
(Fig.~\ref{SOC}b). However, we also observe supercritical bumps for large
$s$ and $N$, meaning that the network is in a slightly supercritical state.

This empirical evidence is supported by a mean-field analysis of 
equation~(\ref{gt}). 
Averaging over the sites, we have for the average gain: 
\begin{equation}
\Gamma[t+1] = \Gamma[t] +\frac{1}{\tau}\left( A - \Gamma[t] \right) 
- u \rho[t] \Gamma[t] \:.
\end{equation}
In the stationary state, we have $\Gamma[t+1] = \Gamma[t] = \Gamma^*$, so:
\begin{equation}
\left(\frac{1}{\tau}+ u \rho^* \right) \Gamma^* =\frac{A}{\tau}\:.
\end{equation}
But we have the relation 
\begin{equation}
\rho^* = C(\Gamma^* - \gcrit)/\Gamma^*
\label{rhostar}
\end{equation}
near the critical region, where $C$ is a constant 
that depends on $\Phi(V)$ and $\mu$, for example, with $\mu=0$,
$C=1$ for $\Phi$ linear monomial model. So:
\begin{equation}
\left(\frac{\Gamma^*}{\tau}+ u C \Gamma^* - u C \gcrit \right) \Gamma^* 
=\frac{A \Gamma^* }{\tau}\:.
\end{equation}
Eliminating the common factor $\Gamma^*$, and dividing by $uC$, we have:
\begin{equation}
\left(1+ \frac{1}{u C \tau}\right) \Gamma^* 
= \gcrit +\frac{ A}{u C \tau} \:.
\end{equation}
Now, call $x = 1/(u C \tau)$. Then, we have:
\begin{equation}
\Gamma^* = \frac{\gcrit + A x}{1+x}\:.
\end{equation}
The fine tuning solution is to put by hand $A = \gcrit$, 
which leads to $\Gamma^* = \gcrit$
independent of $x$. This fine tuning solution should not 
be allowed in a true SOC scenario. 
So, suppose that $A = B \gcrit$. Then, we have:
\begin{equation}
\Gamma^* = \gcrit \frac{1+Bx}{1+x}\:.
\end{equation}
Now we see that to have a critical or supercritical state 
(where equation~(\ref{rhostar}) holds) we must 
have $B > 1$, otherwise we fall in the
subcritical state $\Gamma^* < \gcrit $ where $\rho^* = 0$ and our
mean-field calculation  is not valid. 
A first order approximation leads to:
\begin{equation}
\Gamma^* = \gcrit + (A-\gcrit) x + O(x^2) \:.
\end{equation}
This mean-field calculation shows that, if $x \rightarrow 0$, we obtain
a SOC state $\Gamma^* \rightarrow \gcrit$. However, 
the strict case $x \rightarrow 0$ would require a scaling 
$\tau = O(N^a)$ with an exponent $a>0$, as done previously for dynamic 
synapses~\cite{Levina2007,Bonachela2010,Costa2015,Campos2016}.

However, if we want to avoid the non-biological 
scaling $\tau(N) = O(N^a)$, we can use biologically 
reasonable parameters like 
$\tau \in [10,1000]$ ms, $u = [0.1,1]$, 
$C=1$ and $A \in [1.1,2] \gcrit$. In particular, if 
$\tau = 1000, u=1$ and $A=1.1$, we have $x = 0.001$ and:
\begin{equation}
\Gamma^* \approx 1.0001 \gcrit \:. 
\end{equation}
Even a more conservative value $\tau = 100$ ms gives 
$\Gamma^* \approx 1.001 \gcrit$. Although 
not perfect SOC~\cite{Markovic2014}, 
this result is totally sufficient to explain 
power law neuronal avalanches.
We call this phenomena self-organized supercriticality (SOSC),
where the supercriticality can be very small.
We must yet determine the volume of parameter space $(\tau, A, u)$
where the SOSC phenomenon holds. In the case of 
dynamic synapses $W_{ij}[t]$, this parametric 
volume is very large~\cite{Costa2015,Campos2016} and we conjecture 
that the same occurs for the dynamic gains $\Gamma_i[t]$.
This shall be studied in detail in another paper.

\section*{Discussion}

{\bf Stochastic model:}
The stochastic neuron introduced by Galves and 
Löcherbach~\cite{Galves2013,Galves2016} is an interesting
element for studies of networks of spiking neurons because
it enables exact analytic results and simple numerical calculations. 
While the LSIF models of Soula \emph{et al.}~\cite{Soula2006} and 
Cessac~\cite{Cessac2008,Cessac2010,Cessac2011} introduce 
stochasticity in the neuron's behavior by adding noise terms 
to its potential, the GL model is agnostic about the origin
of noise and randomness (which can be a good thing when several noise
sources are present). All the random behavior is grouped at the
single firing function $\Phi(V)$.

{\bf Phase transitions:}
Networks of GL neurons display a variety of dynamical 
states with interesting phase transitions. 
We looked for stationary regimes in 
such networks, for some specific firing functions 
$\Phi(V)$ with no spontaneous activity 
at the baseline potential (that is, with $\Phi(0) = 0$ and $I=0$).  
We studied the changes in those regimes as a function of the mean 
synaptic weight $W$ and mean neuronal gain $\Gamma$. We found basically 
tree kinds of phase transition, depending of the behavior 
of $\Phi(V) \propto V^r$ for low $V$:
\begin{itemize}
\item[] $r < 1$: A ceaseless dynamic regime with no phase 
transitions ($\Wcrit=0$) similar to that found by Larremore \emph{et al.}~\cite{Larremore2014};
\item[] $r = 1$: A continuous (second order) absorbing 
state phase transition in the Directed Percolation 
universality class usual in SOC models~\cite{Chialvo2010,Markovic2014,Hesse2015,Costa2015,Campos2016};
\item[] $r > 1$: Discontinuous (first order) absorbing state transitions. 
\end{itemize}
We also observed discontinuous phase transitions for any $r>0$ when
the neurons have a firing threshold $\Vfire > 0$.

The deterministic LIF neuron models, which do not have noise, do not seem 
to allow these kinds of transitions~\cite{Brette2007,Ostojic2014,Torres2015}.
The model studied by Larremore \emph{et al.}~\cite{Larremore2014} is equivalent
to the GL model with monomial saturating firing function with 
$r=1, \Vfire = 0, \mu=0$ and $\Gamma = 1$.  
They did not report any phase transition (perhaps because of the effect of
inhibitory neurons in their network), but found a
ceaseless activity very similar to what we observed with $r < 1$.

{\bf Avalanches:}
In the case of second-order phase transitions 
($\Phi(0)=0, r = 1, \Vfire=0$),
we detected firing avalanches at the critical boundary $\gcrit = 1/W$ 
whose size and duration power law distributions present
the standard mean-field exponents $\tau_S = 3/2$ and $\tau_D = 2$. 
We observed a very good finite-scaling and data collapse behavior, 
with finite-size exponents $c_S = 1$ and $c_D=1/2$.

{\bf Maximal susceptibility and optimal dynamic range at criticality:}
Maximal susceptibility means maximal sensitivity to inputs, in special to
weak inputs, which seems to be an interesting property in biological terms.
So, this is a new example of optimization of information processing 
at criticality. We also observed, for small $I$, the behavior 
$\rho(I) \propto I^m$ with a fractionary Stevens's exponent
$m = 1/\delta= 1/2$. Fractionary Stevens's exponents maximize 
the network dynamic range since,
outside criticality, we have only a input-output proportional behavior 
$\rho(I)\propto I$~\cite{Kinouchi2006}. As an example, 
in non-critical systems, an input
range of $1-10000$ spikes/s, arriving to the neurons due to their
extensive dendritic arbors, must be mapped onto a range also of 
$1-10000$ spikes/s in each neuron, which is biologically impossible
because neuronal firing do not span four orders of magnitude. 
However, at criticality, since $\rho(I) \propto I^{1/\delta} = \sqrt{I}$, 
a similar input range needs to be mapped only to an output range of $1-100$
spikes/s, which is biologically possible.
Optimal dynamic range and maximal susceptibility to small inputs
constitute prime biological motivations
to neuronal networks self-organize toward criticality.

{\bf Self-organized criticality:}
One way to achieve this goal is to use dynamical synapses $W_{ij}[t]$, 
in a way that mimics the loss of strength after a synaptic discharge 
(presumably due to neurotransmitter vesicles depletion), 
and the subsequent slow recovery~\cite{Levina2007,Bonachela2010,
Costa2015,Campos2016}: 
\begin{equation}
 W_{ij}[t+1] = W_{ij}[t] +\frac{1}{\tau}\left(A - W_{ij}[t]\right) 
 - u  W_{ij}[t] X_j[t]  \:.
\label{Wt}
\end{equation} 
The parameters are 
the synaptic recovery time $\tau$, the asymptotic value $A$, and  
the fraction $u$ of synaptic weight lost after firing. This synaptic dynamics 
has been examined 
in~\cite{Levina2007,Bonachela2010,Costa2015,Campos2016}.
For our all-to-all coupled network, 
we have $K = N-1$ and $N(N-1)$ dynamic equations for the $W_{ij}s$.
This is a huge number, for example $O(10^8)$ equations, 
even for a moderate network of $N=10^4$ neurons~\cite{Costa2015,Campos2016}.
The possibility of well behaved 
SOC in bulk dissipative systems with loading  is discussed 
in~\cite{Bonachela2009,Bonachela2010,Markovic2014}. Further considerations
for systems with conservation on the average at the stationary state, 
as occurs in our model, are made in \cite{Costa2015,Campos2016}.

Inspired by the presence of the critical boundary, 
we proposed a new mechanism for 
short-scale neural network plasticity, based on dynamic
neuron gains $\Gamma_i[t]$ instead of the 
above dynamic synaptic weights. 
This new mechanism is biologically plausible, probably related
an activity-dependent firing probability 
at the axon initial segment (AIS)~\cite{Kole2012,Platkiewicz2010}, 
and was found to be sufficient to self-organize the network
near the critical region.
We obtained good data collapse and finite-size behavior for the
$P_S(S)$ distributions but,
in contrast with the static model, we get a finite-size exponent 
$c_S=2/3$. The reason for this difference is not clear by now, but
we notice that such $c_S=2/3$ exponent has been found previously 
in the Pruessner–Jensen SOC model and explained by
a field theory elaborated for such systems~\cite{Bonachela2009}.

The great advantage of this new SOC mechanism 
is its computational efficiency: when 
simulating $N$ neurons with $K$ synapses each, there are only $N$ dynamic 
equations for the gains $\Gamma_i[t]$, instead of $NK$ equations
for the synaptic weights $W_{ij}[t]$. Notice that, for the all-to-all
coupling network studied here, this means $O(N^2)$ equations for
dynamic synapse but only $O(N)$ equations for dynamic gains. This makes
a huge difference for the network sizes that can be simulated.

We stress that, since we used $\tau$ finite, the criticality is not
perfect ($\Gamma^*/\gcrit  \in [1.001;1.01] $). 
So, we called it a self-organized super-criticality (SOSC) phenomenon. 
Interestingly, SOSC would be a concretization of Turing's intuition that the
best brain operating point is slightly supercritical~\cite{Turing1950}.

We speculate that this slightly supercriticality could explain 
why humans are so prone to supercritical-like pathological states like
epilepsy~\cite{Hesse2015} (prevalence $1.7 \%$) and mania
(prevalence $2.6 \%$ in the population). Our mechanism suggests that such
pathological states arises from small gain depression $u$
or small gain recovery time $\tau$. These parameters are experimentally
related to firing rate adaptation and perhaps
our proposal could be experimentally studied 
in normal and pathological tissues.

We also conjecture that this supecriticality in the whole network
could explain the \emph{Subsamplig Paradox} in neuronal avalanches:
since the initial experimental protocols~\cite{Beggs2003,Markovic2014}, 
critical power laws have been seem when using arrays of $N_e = 32 - 512$
electrodes, which are a very small numbers
compared to the full biological network size with $N = O(10^6-10^9)$ neurons.
This situation $N_e << N$ has been called 
\emph{subsampling}~\cite{Priesemann2009,Ribeiro2010,Ribeiro2014}. 

The paradox occurs because models that present 
good power laws for avalanches
measured over the total number of neurons $N$, under subsampling 
present only exponential tails or log-normal behaviors\cite{Ribeiro2014}.
No model, to the best of our knowledge, 
has solved this paradox~\cite{Markovic2014}.
Our dynamic gains, which produce supercritical 
states like $\Gamma^* = 1.01\gcrit$, 
could be a solution to the paradox if the supercriticality in the whole
network, described by a power law with a supercritical bump
for large avalanches, turns out to be described by an apparent pure power law under
subsampling. This possibility will be fully explored in another paper.

{\bf Directions for future research:}
Future research could investigate other
network topologies and firing functions, heterogeneous
networks, the effect of inhibitory neurons~\cite{Larremore2014,Ostojic2014}, 
and network learning. The study of 
self-organized supercriticality (and subsampling) 
with GL neurons and dynamic neuron gains
is particularly promising.

\section*{Methods}

{\bf Numerical Calculations:} All numerical calculations are done by using
MATLAB software.
{\bf Simulation procedures:} Simulation codes are 
made in Fortran90 and C++11.
The avalanche statistics were obtained by simulating the evolution of 
finite networks of $N$ neurons, with uniform synaptic strengths 
$W_{ij} = W$ ($W_{ii} = 0$), $\Phi(V)$ monomial linear ($r=1$) 
and critical parameter values $\Wcrit = 1$ and $\gcrit = 1$.  
Each avalanche was started with all neuron potentials 
$V_i[0] = \Vreset=0$ and forcing the firing of 
a single random neuron $i$ by setting $X_i[0]= 1$. 

In contrast to standard integrate-and fire~\cite{Levina2007,Bonachela2010} 
or automata networks~\cite{Kinouchi2006,Costa2015,Campos2016}, stochastic
networks can fire even after intervals with no firing ($\rho[t]=0$) because
membrane voltages $V_[t]$ are not necessarily zero and $\Phi(V)$ can
produce new delayed firings. So, our criteria to define avalanches 
is slightly different from previous literature:
the network was simulated according to 
equation~(\ref{modeldiscrete}) until all potentials 
had decayed to such low values that $\sum_i^N V_i[t] < 10^{-20}$, so 
further spontaneous firing would not be expected to occur
for thousands of steps, which defines a \emph{stop time}. 
Then, the total number of firings $s$ is counted from the 
first firing up to this stop time.

The correct finite-size scaling for avalanche duration is
obtained by defining the duration as $d = d_{bare} + 5$ time steps,
where $d_{bare}$ is the measured duration in the simulation. These 
extra five time steps probably arise from the new definition of
avalanche used for these stochastic neurons.

\bibliography{bib_abrev}

\begin{thebibliography}{10}
\expandafter\ifx\csname url\endcsname\relax
  \def\url#1{\texttt{#1}}\fi
\expandafter\ifx\csname urlprefix\endcsname\relax\def\urlprefix{URL }\fi
\providecommand{\bibinfo}[2]{#2}
\providecommand{\eprint}[2][]{\url{#2}}

\bibitem{Turing1950}
\bibinfo{author}{Turing, A.~M.}
\newblock \bibinfo{title}{Computing machinery and intelligence.}
\newblock \emph{\bibinfo{journal}{Mind}} \textbf{\bibinfo{volume}{59}},
  \bibinfo{pages}{433--460} (\bibinfo{year}{1950}).

\bibitem{Chialvo2010}
\bibinfo{author}{Chialvo, D.~R.}
\newblock \bibinfo{title}{Emergent complex neural dynamics}.
\newblock \emph{\bibinfo{journal}{Nat. Phys.}} \textbf{\bibinfo{volume}{6}},
  \bibinfo{pages}{744--750} (\bibinfo{year}{2010}).

\bibitem{Hesse2015}
\bibinfo{author}{Hesse, J.} \& \bibinfo{author}{Gross, T.}
\newblock \bibinfo{title}{Self-organized criticality as a fundamental property
  of neural systems}.
\newblock \emph{\bibinfo{journal}{Front. Syst. Neurosci.}}
  (\bibinfo{year}{2015}).

\bibitem{Kinouchi2006}
\bibinfo{author}{Kinouchi, O.} \& \bibinfo{author}{Copelli, M.}
\newblock \bibinfo{title}{Optimal dynamical range of excitable networks at
  criticality}.
\newblock \emph{\bibinfo{journal}{Nat. Phys.}} \textbf{\bibinfo{volume}{2}},
  \bibinfo{pages}{348--351} (\bibinfo{year}{2006}).

\bibitem{Beggs2008}
\bibinfo{author}{Beggs, J.~M.}
\newblock \bibinfo{title}{The criticality hypothesis: how local cortical
  networks might optimize information processing}.
\newblock \emph{\bibinfo{journal}{Philos. Trans. R. Soc. A}}
  \textbf{\bibinfo{volume}{366}}, \bibinfo{pages}{329--343}
  (\bibinfo{year}{2008}).

\bibitem{Shew2009}
\bibinfo{author}{Shew, W.~L.}, \bibinfo{author}{Yang, H.},
  \bibinfo{author}{Petermann, T.}, \bibinfo{author}{Roy, R.} \&
  \bibinfo{author}{Plenz, D.}
\newblock \bibinfo{title}{Neuronal avalanches imply maximum dynamic range in
  cortical networks at criticality}.
\newblock \emph{\bibinfo{journal}{J. Neurosci.}} \textbf{\bibinfo{volume}{29}},
  \bibinfo{pages}{15595--15600} (\bibinfo{year}{2009}).

\bibitem{Massobrio2015}
\bibinfo{author}{Massobrio, P.}, \bibinfo{author}{de~Arcangelis, L.},
  \bibinfo{author}{Pasquale, V.}, \bibinfo{author}{Jensen, H.~J.} \&
  \bibinfo{author}{Plenz, D.}
\newblock \bibinfo{title}{Criticality as a signature of healthy neural
  systems}.
\newblock \emph{\bibinfo{journal}{Front. Syst. Neurosci.}}
  \textbf{\bibinfo{volume}{9}} (\bibinfo{year}{2015}).

\bibitem{Herz1995}
\bibinfo{author}{Herz, A.~V.} \& \bibinfo{author}{Hopfield, J.~J.}
\newblock \bibinfo{title}{Earthquake cycles and neural reverberations:
  collective oscillations in systems with pulse-coupled threshold elements}.
\newblock \emph{\bibinfo{journal}{Phys. Rev. Lett.}}
  \textbf{\bibinfo{volume}{75}}, \bibinfo{pages}{1222} (\bibinfo{year}{1995}).

\bibitem{Beggs2003}
\bibinfo{author}{Beggs, J.~M.} \& \bibinfo{author}{Plenz, D.}
\newblock \bibinfo{title}{Neuronal avalanches in neocortical circuits}.
\newblock \emph{\bibinfo{journal}{J. Neurosci.}} \textbf{\bibinfo{volume}{23}},
  \bibinfo{pages}{11167--11177} (\bibinfo{year}{2003}).

\bibitem{Markovic2014}
\bibinfo{author}{Markovi{\'c}, D.} \& \bibinfo{author}{Gros, C.}
\newblock \bibinfo{title}{Power laws and self-organized criticality in theory
  and nature}.
\newblock \emph{\bibinfo{journal}{Phys. Rep.}} \textbf{\bibinfo{volume}{536}},
  \bibinfo{pages}{41--74} (\bibinfo{year}{2014}).

\bibitem{Arcangelis2006}
\bibinfo{author}{de~Arcangelis, L.}, \bibinfo{author}{Perrone-Capano, C.} \&
  \bibinfo{author}{Herrmann, H.~J.}
\newblock \bibinfo{title}{Self-organized criticality model for brain
  plasticity}.
\newblock \emph{\bibinfo{journal}{Phys. Rev. Lett.}}
  \textbf{\bibinfo{volume}{96}}, \bibinfo{pages}{028107}
  (\bibinfo{year}{2006}).

\bibitem{Levina2007}
\bibinfo{author}{Levina, A.}, \bibinfo{author}{Herrmann, J.~M.} \&
  \bibinfo{author}{Geisel, T.}
\newblock \bibinfo{title}{Dynamical synapses causing self-organized criticality
  in neural networks}.
\newblock \emph{\bibinfo{journal}{Nat. Phys.}} \textbf{\bibinfo{volume}{3}},
  \bibinfo{pages}{857--860} (\bibinfo{year}{2007}).

\bibitem{Bonachela2010}
\bibinfo{author}{Bonachela, J.~A.}, \bibinfo{author}{De~Franciscis, S.},
  \bibinfo{author}{Torres, J.~J.} \& \bibinfo{author}{Mu{\~n}oz, M.~A.}
\newblock \bibinfo{title}{Self-organization without conservation: are neuronal
  avalanches generically critical?}
\newblock \emph{\bibinfo{journal}{J. Stat. Mech. -Theory Exp.}}
  \textbf{\bibinfo{volume}{2010}}, \bibinfo{pages}{P02015}
  (\bibinfo{year}{2010}).

\bibitem{Arcangelis2012}
\bibinfo{author}{De~Arcangelis, L.}
\newblock \bibinfo{title}{Are dragon-king neuronal avalanches dungeons for
  self-organized brain activity?}
\newblock \emph{\bibinfo{journal}{Eur. Phys. J. Spec. Top.}}
  \textbf{\bibinfo{volume}{205}}, \bibinfo{pages}{243--257}
  (\bibinfo{year}{2012}).

\bibitem{Costa2015}
\bibinfo{author}{Costa, A.}, \bibinfo{author}{Copelli, M.} \&
  \bibinfo{author}{Kinouchi, O.}
\newblock \bibinfo{title}{Can dynamical synapses produce true self-organized
  criticality?}
\newblock \emph{\bibinfo{journal}{J. Stat. Mech. -Theory Exp.}}
  \textbf{\bibinfo{volume}{2015}}, \bibinfo{pages}{P06004}
  (\bibinfo{year}{2015}).

\bibitem{Kessenich2016}
\bibinfo{author}{van Kessenich, L.~M.}, \bibinfo{author}{de~Arcangelis, L.} \&
  \bibinfo{author}{Herrmann, H.}
\newblock \bibinfo{title}{Synaptic plasticity and neuronal refractory time
  cause scaling behaviour of neuronal avalanches}.
\newblock \emph{\bibinfo{journal}{Sci. Rep.}} \textbf{\bibinfo{volume}{6}}
  (\bibinfo{year}{2016}).

\bibitem{Campos2016}
\bibinfo{author}{Campos, J.}, \bibinfo{author}{Costa, A.},
  \bibinfo{author}{Copelli, M.} \& \bibinfo{author}{Kinouchi, O.}
\newblock \bibinfo{title}{Correlations induced by depressing synapses in
  quenched critically self-organized networks}.
\newblock \emph{\bibinfo{journal}{arXiv:1604.05779}}  (\bibinfo{year}{2016}).
\newblock \bibinfo{note}{(Submmited to Phys. Rev. E)}.

\bibitem{Ermentrout2001}
\bibinfo{author}{Ermentrout, B.}, \bibinfo{author}{Pascal, M.} \&
  \bibinfo{author}{Gutkin, B.}
\newblock \bibinfo{title}{The effects of spike frequency adaptation and
  negative feedback on the synchronization of neural oscillators}.
\newblock \emph{\bibinfo{journal}{Neural Comput.}}
  \textbf{\bibinfo{volume}{13}}, \bibinfo{pages}{1285--1310}
  (\bibinfo{year}{2001}).

\bibitem{Benda2003}
\bibinfo{author}{Benda, J.} \& \bibinfo{author}{Herz, A.~V.}
\newblock \bibinfo{title}{A universal model for spike-frequency adaptation}.
\newblock \emph{\bibinfo{journal}{Neural Comput.}}
  \textbf{\bibinfo{volume}{15}}, \bibinfo{pages}{2523--2564}
  (\bibinfo{year}{2003}).

\bibitem{Buonocore2016}
\bibinfo{author}{Buonocore, A.}, \bibinfo{author}{Caputo, L.},
  \bibinfo{author}{Pirozzi, E.} \& \bibinfo{author}{Carfora, M.~F.}
\newblock \bibinfo{title}{A leaky integrate-and-fire model with adaptation for
  the generation of a spike train.}
\newblock \emph{\bibinfo{journal}{Math. Biosci. Eng.}}
  \textbf{\bibinfo{volume}{13}}, \bibinfo{pages}{483--493}
  (\bibinfo{year}{2016}).

\bibitem{Galves2013}
\bibinfo{author}{Galves, A.} \& \bibinfo{author}{L{\"o}cherbach, E.}
\newblock \bibinfo{title}{Infinite systems of interacting chains with memory of
  variable length — a stochastic model for biological neural nets}.
\newblock \emph{\bibinfo{journal}{J. Stat. Phys.}}
  \textbf{\bibinfo{volume}{151}}, \bibinfo{pages}{896--921}
  (\bibinfo{year}{2013}).

\bibitem{Lapicque1907}
\bibinfo{author}{Lapicque, L.}
\newblock \bibinfo{title}{Recherches quantitatives sur l’excitation
  électrique des nerfs traitée comme une polarisation}.
\newblock \emph{\bibinfo{journal}{J. Physiol. Pathol. Gen.}}
  \textbf{\bibinfo{volume}{9}}, \bibinfo{pages}{620–635}
  (\bibinfo{year}{1907}).
\newblock \bibinfo{note}{Translation: Brunel, N. {\&} van Rossum, M.C.
  Quantitative investigations of electrical nerve excitation treated as
  polarization. Biol. Cybernetics {\bf 97}, 341--349 (2007).}

\bibitem{Gerstein1964}
\bibinfo{author}{Gerstein, G.~L.} \& \bibinfo{author}{Mandelbrot, B.}
\newblock \bibinfo{title}{Random walk models for the spike activity of a single
  neuron}.
\newblock \emph{\bibinfo{journal}{Biophys. J.}} \textbf{\bibinfo{volume}{4}},
  \bibinfo{pages}{41} (\bibinfo{year}{1964}).

\bibitem{Burkitt2006a}
\bibinfo{author}{Burkitt, A.~N.}
\newblock \bibinfo{title}{A review of the integrate-and-fire neuron model: I.
  homogeneous synaptic input}.
\newblock \emph{\bibinfo{journal}{Biol. Cybern.}}
  \textbf{\bibinfo{volume}{95}}, \bibinfo{pages}{1--19} (\bibinfo{year}{2006}).

\bibitem{Burkitt2006b}
\bibinfo{author}{Burkitt, A.~N.}
\newblock \bibinfo{title}{A review of the integrate-and-fire neuron model:
  {II}. inhomogeneous synaptic input and network properties}.
\newblock \emph{\bibinfo{journal}{Biol. Cybern.}}
  \textbf{\bibinfo{volume}{95}}, \bibinfo{pages}{97--112}
  (\bibinfo{year}{2006}).

\bibitem{Naud2012}
\bibinfo{author}{Naud, R.} \& \bibinfo{author}{Gerstner, W.}
\newblock \bibinfo{title}{The performance (and limits) of simple neuron models:
  generalizations of the leaky integrate-and-fire model}.
\newblock In \emph{\bibinfo{booktitle}{Computational Systems Neurobiology}},
  \bibinfo{pages}{163--192} (\bibinfo{publisher}{Springer},
  \bibinfo{year}{2012}).

\bibitem{Brette2007}
\bibinfo{author}{Brette, R.} \emph{et~al.}
\newblock \bibinfo{title}{Simulation of networks of spiking neurons: a review
  of tools and strategies}.
\newblock \emph{\bibinfo{journal}{J. Comput. Neurosci.}}
  \textbf{\bibinfo{volume}{23}}, \bibinfo{pages}{349--398}
  (\bibinfo{year}{2007}).

\bibitem{Brette2015}
\bibinfo{author}{Brette, R.}
\newblock \bibinfo{title}{What is the most realistic single-compartment model
  of spike initiation?}
\newblock \emph{\bibinfo{journal}{PLoS Comput. Biol.}}
  \textbf{\bibinfo{volume}{11}}, \bibinfo{pages}{e1004114}
  (\bibinfo{year}{2015}).

\bibitem{Benayoun2010}
\bibinfo{author}{Benayoun, M.}, \bibinfo{author}{Cowan, J.~D.},
  \bibinfo{author}{van Drongelen, W.} \& \bibinfo{author}{Wallace, E.}
\newblock \bibinfo{title}{Avalanches in a stochastic model of spiking neurons}.
\newblock \emph{\bibinfo{journal}{PLoS Comput. Biol.}}
  \textbf{\bibinfo{volume}{6}}, \bibinfo{pages}{e1000846}
  (\bibinfo{year}{2010}).

\bibitem{Ostojic2014}
\bibinfo{author}{Ostojic, S.}
\newblock \bibinfo{title}{Two types of asynchronous activity in networks of
  excitatory and inhibitory spiking neurons}.
\newblock \emph{\bibinfo{journal}{Nat. Neurosci.}}
  \textbf{\bibinfo{volume}{17}}, \bibinfo{pages}{594--600}
  (\bibinfo{year}{2014}).

\bibitem{Torres2015}
\bibinfo{author}{Torres, J.~J.} \& \bibinfo{author}{Marro, J.}
\newblock \bibinfo{title}{Brain performance versus phase transitions}.
\newblock \emph{\bibinfo{journal}{Sci. Rep.}} \textbf{\bibinfo{volume}{5}}
  (\bibinfo{year}{2015}).

\bibitem{Platkiewicz2010}
\bibinfo{author}{Platkiewicz, J.} \& \bibinfo{author}{Brette, R.}
\newblock \bibinfo{title}{A threshold equation for action potential
  initiation}.
\newblock \emph{\bibinfo{journal}{PLoS Comput. Biol.}}
  \textbf{\bibinfo{volume}{6}}, \bibinfo{pages}{e1000850}
  (\bibinfo{year}{2010}).

\bibitem{Mcdonnell2016}
\bibinfo{author}{McDonnell, M.~D.}, \bibinfo{author}{Goldwyn, J.~H.} \&
  \bibinfo{author}{Lindner, B.}
\newblock \bibinfo{title}{Editorial: Neuronal stochastic variability:
  Influences on spiking dynamics and network activity}.
\newblock \emph{\bibinfo{journal}{Front. Comput. Neurosci.}}
  \textbf{\bibinfo{volume}{10}} (\bibinfo{year}{2016}).

\bibitem{Soula2006}
\bibinfo{author}{Soula, H.}, \bibinfo{author}{Beslon, G.} \&
  \bibinfo{author}{Mazet, O.}
\newblock \bibinfo{title}{Spontaneous dynamics of asymmetric random recurrent
  spiking neural networks}.
\newblock \emph{\bibinfo{journal}{Neural Comput.}}
  \textbf{\bibinfo{volume}{18}}, \bibinfo{pages}{60--79}
  (\bibinfo{year}{2006}).

\bibitem{Cessac2008}
\bibinfo{author}{Cessac, B.}
\newblock \bibinfo{title}{A discrete time neural network model with spiking
  neurons}.
\newblock \emph{\bibinfo{journal}{J Math Biol.}} \textbf{\bibinfo{volume}{56}},
  \bibinfo{pages}{311--345} (\bibinfo{year}{2008}).

\bibitem{Cessac2010}
\bibinfo{author}{Cessac, B.}
\newblock \bibinfo{title}{A view of neural networks as dynamical systems}.
\newblock \emph{\bibinfo{journal}{Int. J. Bifurcation Chaos}}
  \textbf{\bibinfo{volume}{20}}, \bibinfo{pages}{1585--1629}
  (\bibinfo{year}{2010}).

\bibitem{Cessac2011}
\bibinfo{author}{Cessac, B.}
\newblock \bibinfo{title}{A discrete time neural network model with spiking
  neurons: {II} : Dynamics with noise}.
\newblock \emph{\bibinfo{journal}{J Math Biol.}} \textbf{\bibinfo{volume}{62}},
  \bibinfo{pages}{863--900} (\bibinfo{year}{2011}).

\bibitem{DeMasi2015}
\bibinfo{author}{De~Masi, A.}, \bibinfo{author}{Galves, A.},
  \bibinfo{author}{L{\"o}cherbach, E.} \& \bibinfo{author}{Presutti, E.}
\newblock \bibinfo{title}{Hydrodynamic limit for interacting neurons}.
\newblock \emph{\bibinfo{journal}{J. Stat. Phys.}}
  \textbf{\bibinfo{volume}{158}}, \bibinfo{pages}{866--902}
  (\bibinfo{year}{2015}).

\bibitem{Duarte2014}
\bibinfo{author}{Duarte, A.} \& \bibinfo{author}{Ost, G.}
\newblock \bibinfo{title}{A model for neural activity in the absence of
  external stimuli}.
\newblock \emph{\bibinfo{journal}{Markov Process. Relat. Fields}}
  \textbf{\bibinfo{volume}{22}}, \bibinfo{pages}{37--52}
  (\bibinfo{year}{2016}).

\bibitem{Duarte2015}
\bibinfo{author}{Duarte, A.}, \bibinfo{author}{Ost, G.} \&
  \bibinfo{author}{Rodr{\'\i}guez, A.~A.}
\newblock \bibinfo{title}{Hydrodynamic limit for spatially structured
  interacting neurons}.
\newblock \emph{\bibinfo{journal}{J. Stat. Phys.}}
  \textbf{\bibinfo{volume}{161}}, \bibinfo{pages}{1163--1202}
  (\bibinfo{year}{2015}).

\bibitem{Galves2016}
\bibinfo{author}{Galves, A.} \& \bibinfo{author}{L{\"o}cherbach, E.}
\newblock \bibinfo{title}{Modeling networks of spiking neurons as interacting
  processes with memory of variable length}.
\newblock \emph{\bibinfo{journal}{J. Soc. Franc. Stat.}}
  \textbf{\bibinfo{volume}{157}}, \bibinfo{pages}{17--32}
  (\bibinfo{year}{2016}).

\bibitem{Larremore2014}
\bibinfo{author}{Larremore, D.~B.}, \bibinfo{author}{Shew, W.~L.},
  \bibinfo{author}{Ott, E.}, \bibinfo{author}{Sorrentino, F.} \&
  \bibinfo{author}{Restrepo, J.~G.}
\newblock \bibinfo{title}{Inhibition causes ceaseless dynamics in networks of
  excitable nodes}.
\newblock \emph{\bibinfo{journal}{Phys. Rev. Lett.}}
  \textbf{\bibinfo{volume}{112}}, \bibinfo{pages}{138103}
  (\bibinfo{year}{2014}).

\bibitem{Virkar2016}
\bibinfo{author}{Virkar, Y.~S.}, \bibinfo{author}{Shew, W.~L.},
  \bibinfo{author}{Restrepo, J.~G.} \& \bibinfo{author}{Ott, E.}
\newblock \bibinfo{title}{Metabolite transport through glial networks
  stabilizes the dynamics of learning}.
\newblock \emph{\bibinfo{journal}{arXiv:1605.03090}}  (\bibinfo{year}{2016}).

\bibitem{Cooper2005}
\bibinfo{author}{Cooper, S.~J.}
\newblock \bibinfo{title}{Donald o. hebb's synapse and learning rule: a history
  and commentary}.
\newblock \emph{\bibinfo{journal}{Neurosci. Biobehav. Rev.}}
  \textbf{\bibinfo{volume}{28}}, \bibinfo{pages}{851--874}
  (\bibinfo{year}{2005}).

\bibitem{Tsodyks1998}
\bibinfo{author}{Tsodyks, M.}, \bibinfo{author}{Pawelzik, K.} \&
  \bibinfo{author}{Markram, H.}
\newblock \bibinfo{title}{Neural networks with dynamic synapses}.
\newblock \emph{\bibinfo{journal}{Neural Comput.}}
  \textbf{\bibinfo{volume}{10}}, \bibinfo{pages}{821--835}
  (\bibinfo{year}{1998}).

\bibitem{Larremore2011}
\bibinfo{author}{Larremore, D.~B.}, \bibinfo{author}{Shew, W.~L.} \&
  \bibinfo{author}{Restrepo, J.~G.}
\newblock \bibinfo{title}{Predicting criticality and dynamic range in complex
  networks: effects of topology}.
\newblock \emph{\bibinfo{journal}{Phys. Rev. Let.}}
  \textbf{\bibinfo{volume}{106}}, \bibinfo{pages}{058101}
  (\bibinfo{year}{2011}).

\bibitem{Kole2012}
\bibinfo{author}{Kole, M.~H.} \& \bibinfo{author}{Stuart, G.~J.}
\newblock \bibinfo{title}{Signal processing in the axon initial segment}.
\newblock \emph{\bibinfo{journal}{Neuron}} \textbf{\bibinfo{volume}{73}},
  \bibinfo{pages}{235--247} (\bibinfo{year}{2012}).

\bibitem{Lipetz1971}
\bibinfo{author}{Lipetz, L.~E.}
\newblock \bibinfo{title}{The relation of physiological and psychological
  aspects of sensory intensity}.
\newblock In \emph{\bibinfo{booktitle}{Principles of Receptor Physiology}},
  \bibinfo{pages}{191--225} (\bibinfo{publisher}{Springer},
  \bibinfo{year}{1971}).

\bibitem{Naka1966}
\bibinfo{author}{Naka, K.-I.} \& \bibinfo{author}{Rushton, W.~A.}
\newblock \bibinfo{title}{S-potentials from luminosity units in the retina of
  fish (cyprinidae)}.
\newblock \emph{\bibinfo{journal}{J Physiol.}} \textbf{\bibinfo{volume}{185}},
  \bibinfo{pages}{587} (\bibinfo{year}{1966}).

\bibitem{Bonachela2009}
\bibinfo{author}{Bonachela, J.~A.} \& \bibinfo{author}{Mu{\~n}oz, M.~A.}
\newblock \bibinfo{title}{Self-organization without conservation: true or just
  apparent scale-invariance?}
\newblock \emph{\bibinfo{journal}{J. Stat. Mech.-Theory Exp.}}
  \textbf{\bibinfo{volume}{2009}}, \bibinfo{pages}{P09009}
  (\bibinfo{year}{2009}).

\bibitem{Priesemann2009}
\bibinfo{author}{Priesemann, V.}, \bibinfo{author}{Munk, M.~H.} \&
  \bibinfo{author}{Wibral, M.}
\newblock \bibinfo{title}{Subsampling effects in neuronal avalanche
  distributions recorded in vivo}.
\newblock \emph{\bibinfo{journal}{BMC Neurosci.}}
  \textbf{\bibinfo{volume}{10}}, \bibinfo{pages}{40} (\bibinfo{year}{2009}).

\bibitem{Ribeiro2010}
\bibinfo{author}{Ribeiro, T.~L.} \emph{et~al.}
\newblock \bibinfo{title}{Spike avalanches exhibit universal dynamics across
  the sleep-wake cycle}.
\newblock \emph{\bibinfo{journal}{PLoS One}} \textbf{\bibinfo{volume}{5}},
  \bibinfo{pages}{e14129} (\bibinfo{year}{2010}).

\bibitem{Ribeiro2014}
\bibinfo{author}{Ribeiro, T.~L.}, \bibinfo{author}{Ribeiro, S.},
  \bibinfo{author}{Belchior, H.}, \bibinfo{author}{Caixeta, F.} \&
  \bibinfo{author}{Copelli, M.}
\newblock \bibinfo{title}{Undersampled critical branching processes on
  small-world and random networks fail to reproduce the statistics of spike
  avalanches}.
\newblock \emph{\bibinfo{journal}{PLoS One}} \textbf{\bibinfo{volume}{9}},
  \bibinfo{pages}{e94992} (\bibinfo{year}{2014}).

\end{thebibliography}
%\bibliographystyle{ieeetr}
%\bibliographystyle{plain}
%\begin{thebibliography}{10}
%\end{thebibliography}

\section*{Acknowledgements}

This paper results from research activity on the 
FAPESP Center for Neuromathematics (FAPESP grant 2013/07699-0). 
OK and AAC also received support from Núcleo de Apoio à Pesquisa 
CNAIPS-USP and FAPESP (grant 2016/00430-3). 
LB, JS and ACR also received CNPq 
support (grants 165828/2015-3, 310706/2015-7 and 306251/2014-0).
We thank A. Galves for suggestions and revision of the paper, and 
M. Copelli and S. Ribeiro for discussions.

\section*{Author contributions statement}

LB and AAC performed the simulations and prepared all
the figures. OK and JS made the analytic calculations.
OK, JS and LB wrote the paper.
MA and ACR contributed with ideas, the writing of the
paper and citations to the literature.
All authors reviewed the manuscript.

\textbf{Competing financial interests} The authors declare no competing
financial interests. 

\end{document}